\documentclass[sigconf,natbib=true,anonymous=false]{acmart}
\AtBeginDocument{%
  }

\copyrightyear{2025}
\acmYear{2025}
\setcopyright{cc}
\setcctype{by}
\acmConference[ICTIR '25]{Proceedings of the 2025 International ACM SIGIR Conference on Innovative Concepts and Theories in Information Retrieval (ICTIR)}{July 18, 2025}{Padua, Italy}
\acmBooktitle{Proceedings of the 2025 International ACM SIGIR Conference on Innovative Concepts and Theories in Information Retrieval (ICTIR) (ICTIR '25), July 18, 2025, Padua, Italy}
\acmDOI{10.1145/3731120.3744614}
\acmISBN{979-8-4007-1861-8/2025/07}

\begin{document}

\title{The Next Phase of Scientific Fact-Checking: Advanced Evidence Retrieval from Complex Structured Academic Papers}

\author{Xingyu Deng}
\email{xdeng37@sheffield.ac.uk}
\affiliation{%
  \institution{University of Sheffield}
  \city{Sheffield}
  \country{UK}}

\author{Xi Wang}
\email{xi.wang@sheffield.ac.uk}
\affiliation{%
  \institution{University of Sheffield}
  \city{Sheffield}
  \country{UK}}

\author{Mark Stevenson}
\email{mark.stevenson@sheffield.ac.uk}
\affiliation{%
  \institution{University of Sheffield}
  \city{Sheffield}
  \country{UK}}

\renewcommand{\shortauthors}{Xingyu Deng, Xi Wang, Mark Stevenson}

\begin{abstract}
Scientific fact-checking aims to determine the veracity of scientific claims by retrieving and analysing evidence from research literature. The problem is inherently more complex than general fact-checking since it must accommodate the evolving nature of scientific knowledge, the structural complexity of academic literature and the challenges posed by long-form, multimodal scientific expression. However, existing approaches focus on simplified versions of the problem based on small-scale datasets consisting of abstracts rather than full papers, thereby avoiding the distinct challenges associated with processing complete documents. This paper examines the limitations of current scientific fact-checking systems and reveals the many potential features and resources that could be exploited to advance their performance. It identifies key research challenges within evidence retrieval, including (1) evidence-driven retrieval that addresses semantic limitations and topic imbalance (2) time-aware evidence retrieval with citation tracking to mitigate outdated information, (3) structured document parsing to leverage long-range context, (4) handling complex scientific expressions, including tables, figures, and domain-specific terminology and (5) assessing the credibility of scientific literature. Preliminary experiments were conducted to substantiate these challenges and identify potential solutions. This perspective paper aims to advance scientific fact-checking with a specialised IR system tailored for real-world applications.
\end{abstract}





\maketitle

\section{Introduction} 
Fact-checking aims to assess the veracity of factual claims based on credible evidence  \cite{guo2022survey,zeng2021automated} and serves as a crucial safeguard for mitigating misinformation. 
Scientific fact-checking is a specialised variant of this task, grounded in scientific knowledge, with the objective of combating misinformation that affects the public, helping researchers in knowledge discovery and assisting individuals in understanding scientific advancements  \cite{vladika-matthes-2023-scientific}. 
This is particularly important given the rapid emergence of new scientific findings, where both professionals and the public must assess the credibility of information. A prominent case occurred during the COVID-19 pandemic, in which politically motivated misinformation—ranging from inflated infection statistics to unsupported treatments—circulated extensively, eroding public trust and endangering health communication  \cite{luengo2020performance}.
However, existing approaches to scientific fact-checking remain limited, primarily relying on the retrieval of evidence from relatively simple and small-scale sources  \cite{wadden-etal-2020-fact,mohr-etal-2022-covert,wang-etal-2023-check-covid,bulian2020climate,saakyan-etal-2021-covid,sarrouti-etal-2021-evidence-based,kotonya-toni-2020-explainable-automated,wadden-etal-2022-scifact}. For example, SciFact-Open  \cite{wadden-etal-2022-scifact}, the largest available dataset for scientific fact-checking, contains 500,000 documents -- substantially smaller than PubMed, which contains over 37 million biomedical publications. In addition, SciFact-Open consists only of abstracts, rather than full-text papers, thereby excluding critical structural and citation information, ignoring long-range context and scientific expression conveyed through tables and figures.
These design simplifications may hinder the applicability of current approaches in real-world settings, where scientific evidence is embedded in long and structurally complex documents with multimodal content.

Fact-checking is a knowledge-intensive task, where the verification process relies on sourcing evidence from a reliable upstream Information Retrieval (IR) system. Emerging findings indicate the value of effective retrieval in improving fact-checking systems. For example, introducing even a small amount of noise into evidence can significantly degrade fact-checking performance  \cite{sauchuk2022role}. Recent Retrieval-Augmented-Generation (RAG) techniques have been widely used for fact-checking  \cite{dmonte2024claim,raina-gales-2024-question,momii-etal-2024-rag,sevgili-etal-2024-uhh,singal-etal-2024-evidence,ullrich-etal-2024-aic,khaliq-etal-2024-ragar}, where retrieval models are fine-tuned to identify high-quality evidence for claim verification. These observations underscore the critical role of robust evidence retrieval, as an ideal IR system for fact-checking should rank all relevant evidence at the top while filtering out non-evidential noise. Ensuring retrieval robustness is crucial to maintaining sufficient yet relevant evidence, which is essential for improving scientific fact-checking accuracy.

A major trend in fact-checking research is to consider realistic settings that employ rich, diverse and timely evidence sources, as seen in FEVER (using Wikipedia) \cite{thorne-etal-2018-fever} and AVeriTeC (using web-wide resources) \cite{schlichtkrull2024averitec}. In \textbf{document level evidence retrieval}, current general fact-checking systems over-rely on commercial search APIs, which do not consider the specific requirements of fact-checking  \cite{zeng2021automated,thorne-etal-2018-fact,schlichtkrull-etal-2024-automated}.
Such reliance on commercial search APIs -- with limited adaptability -- has left document retrieval methodologies under-explored in fact-checking, especially for domain-specific corpora such as scientific fact-checking.
Current scientific fact-checking systems primarily employ off-the-shelf IR methods  \cite{vladika-matthes-2023-scientific}, such as lexical matching and semantic relevance ranking, which do not scale effectively for large-scale scientific corpora. 
In addition, the distribution of relevant evidence across scientific topics is highly imbalanced, which degrades both retrieval effectiveness and efficiency, especially for claims with scarce supporting literature.
SciFact-Open  \cite{wadden-etal-2022-scifact}, which extends the original SciFact dataset  \cite{wadden-etal-2020-fact} for large-scale evaluation, illustrates this issue: verification performance on SciFact-Open drops by at least 15 F1 points for all well-performed fact-checking systems developed in SciFact  \cite{wadden-etal-2022-scifact}. While increasing corpus size enhances evidence diversity, it also amplifies retrieval noise, reducing efficiency in both retrieval and verification. Beyond that, high semantic relevance does not guarantee high evidential relevance, and irrelevant yet semantically similar documents can introduce noise into downstream verification  \cite{zhang-etal-2023-relevance}. These challenges underscore the necessity of developing tailored document retrieval systems specifically designed for scientific fact-checking, as effective retrieval is a prerequisite for accurate claim verification.

Beyond document-level evidence retrieval, \textbf{within-document evidence retrieval} is also essential for processing complex scientific literature. Scientific papers, unlike general fact-checking documents, are long, structured, domain-specific and involve additional metadata. As scientific fact-checking evolves from abstract-based to full-paper retrieval, retrieval models must account for metadata (e.g., publish date, citations) and complex structured data format (e.g., charts, tables, figures).
This necessitates the adaptation of verification models such as SciBERT  \cite{beltagy-etal-2019-scibert} for domain-specific terminologies  \cite{wadden-etal-2020-fact}, Longformer  \cite{Beltagy2020Longformer} for long-range dependencies  \cite{wadden-etal-2022-multivers} and TAPAS  \cite{herzig-etal-2020-tapas} for tabular data verification  \cite{akhtar-etal-2022-pubhealthtab}. Scientific expressions in academic papers are highly structured and contextually interdependent, where textual content, tabular data, and figures mutually reinforce the conveyed information. However, existing scientific fact-checking systems primarily operate at the abstract level, adopting methodologies similar to general fact-checking ~ \cite{vladika-matthes-2023-scientific,wadden-etal-2020-fact,li2021paragraph,pradeep-etal-2021-scientific,zhang-etal-2021-abstract,wadden-etal-2022-multivers,wuhrl-klinger-2022-entity,kotonya-toni-2020-explainable-automated,sarrouti-etal-2021-evidence-based,saakyan-etal-2021-covid}, albeit incorporating domain-specific models such as BioSentVec  \cite{chen2019biosentvec} and SciBERT  \cite{beltagy-etal-2019-scibert}.
The development of public full-paper datasets aligns with the requirement of real-world scientific fact-checking systems for effective verification. 
This highlights the urgent need for retrieval and verification methodologies that can leverage entire scientific documents. Accordingly, within-document evidence retrieval and its integration into verification pipelines should be explored to fully unlock the potential of scientific literature for fact-checking.

This perspective paper presents a comprehensive examination of the challenges associated with evidence retrieval in scientific fact-checking, \textbf{highlighting challenges that are not typically faced within general fact-checking}, leading to the need for specialised retrieval and verification strategies.
We advocate for proactive research efforts to develop scalable methodologies while addressing the limitations of current datasets. 
We structure our discussion into two parts following a typical fact-checking pipeline: Sections~\ref{sec:beyond_semantics}--\ref{sec:imbalance_resources} explore document-level evidence retrieval while Sections ~\ref{sec:time_citation}--\ref{sec:terminology_complex} explore within-document evidence retrieval in scientific publication for scientific fact-checking.
Each of the following sections identifies a research challenge followed by a tentative and illustrative research direction (\textbf{RD}).

\section{Beyond Semantics}
\label{sec:beyond_semantics}
Evidence retrieval for fact checking is closely related to traditional document retrieval techniques, which typically focus on retrieving documents that are semantically similar to a query or contain matching keywords. While this approach is effective in many scenarios, it often fails to address the ultimate objective of fact-checking -- successful claim verification. Evidence retrieval that relies solely on semantic similarity may prioritise irrelevant or low-context information, introducing noise into the subsequent verification process. 
Recent IR studies  \cite{10.1007/978-3-031-88708-6_3,parry2024context} show that semantic relevance alone may not ensure utility in knowledge-intensive NLP tasks under the RAG framework, suggesting the importance of utility-aware retrieval strategies.
To improve the verification utility of evidence retrieval systems, techniques such as fine-tuning, joint optimisation, and learning from verification feedback have been developed. These approaches leverage relevance labels derived from annotated gold evidence  \cite{pradeep-etal-2021-scientific,zhang-etal-2021-abstract,zheng-etal-2024-evidence,hu2023read,zhang-etal-2023-relevance}. 
Although graded relevance has been extensively explored in general IR, current evidence retrieval systems for fact-checking often oversimplify relevance as binary, failing to differentiate between fully non-evidential and partially relevant evidence.
This coarse-grained labelling scheme fails to differentiate between completely non-evidential documents and partially relevant (plausible) evidence. Negative examples and randomly retrieved examples are equally treated as 0, despite exhibiting varying degrees of evidential support. We argue that evidence retrieval should distinguish between non-evidential information and plausible evidence, enhancing the model's ability to identify previously unobserved but potentially useful evidence within large-scale corpora.

Developing an IR system that can effectively differentiate between evidential and non-evidential information requires access to fine-grained relevance labels during training. However, manually constructing negative samples is both complex and resource-intensive due to the vast number of unlabelled documents and sentences that lack explicit pairing with given claims. Furthermore, assessing the degree of evidential support for a claim within unlabelled documents is inherently challenging. To validate the impact of fine-grained evidential relevance, beyond semantic relevance, we carry out preliminary experiments which explore the use of downstream verification feedback to capture different levels of evidential values.

\begin{figure}[h]
  \centering
  \includegraphics[width=0.8\linewidth]{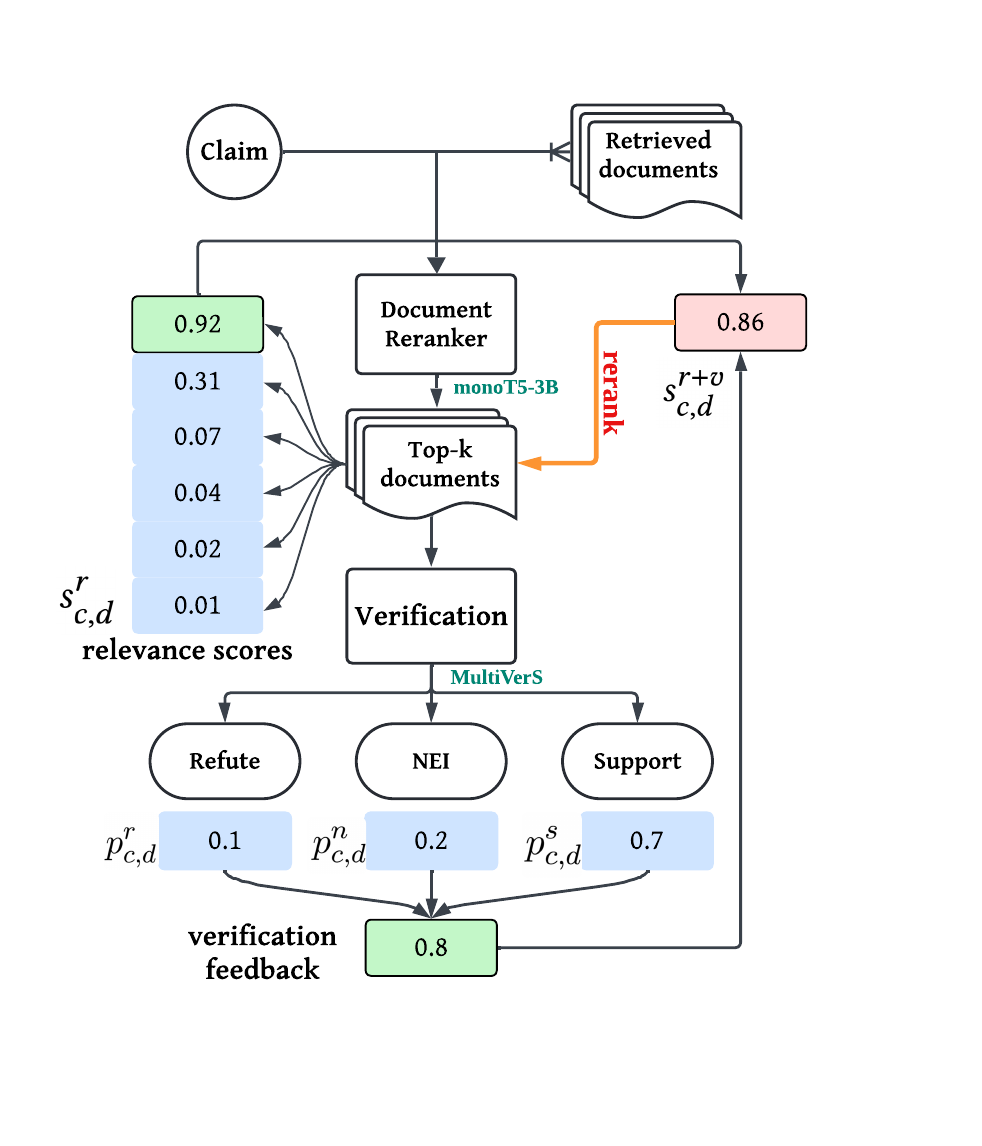}
  \caption{Pipeline of experiment}
  \Description{The technical pipeline for combining relevance score and verification feedback}
\label{fig:feedback_pipeline}
\vspace{-5pt}
\end{figure}

\paragraph{\textbf{Experiment Overview}}
The experiment investigates whether combining verification feedback with semantic relevance improves the performance of document evidence retrieval. Figure~\ref{fig:feedback_pipeline} presents the pipeline, where probabilities from downstream verification serve as feedback. Specifically, we integrate two components: 1) the semantic relevance score, computed using an off-the-shelf reranker model, and (2) the verification success feedback score, derived from a fine-tuned verifier model, indicating the degree to which a document is evidential for a given claim.

\paragraph{\textbf{Approaches}}
\textit{\textbf{monoT5-3B}}  \cite{nogueira-etal-2020-document} has demonstrated strong performance as a reranker for SciFact, as evidenced by its widespread use as a strong baseline in studies  \cite{10.1145/3696410.3714554,sun-etal-2023-chatgpt} on the BEIR benchmark \cite{thakur2021beir}. It also demonstrated state-of-the-art performance in evidence retrieval for verification  \cite{pradeep-etal-2021-scientific,vladika-matthes-2023-scientific,wadden-etal-2022-multivers}. The model assigns a predicted score, $s^{r}_{c,d}$, representing the semantic relevance for a document $d$ to a claim $c$ as defined in Equation~\ref{equa:ir_euqation}.

\begin{equation}
\label{equa:ir_euqation}
  f(c,d) \rightarrow \ s^{r}_{c,d},\ s^{r}_{c,d} \in (0,1) 
\end{equation}

\noindent \textit{\textbf{MultiVerS}} is the best-performing verifier model on SciFact  \cite{wadden-etal-2022-multivers,vladika-matthes-2023-scientific}. We reproduced this model using the official implementation\footnote{\url{https://github.com/dwadden/multivers}} while adjusting the negative sampling parameter from 20 to 5 to avoid over-fitted verification feedback. The model predicts the verification outcome as per the calculated probabilities for a document $d$ either supporting $p^{r}_{c,d}$, refuting $p^{s}_{c,d}$ or providing insufficient information $p^{n}_{c,d}$, relative to a claim $c$, as follows:
\begin{equation}
\label{equa:verifier_equation}
V(c,d) = argmax(p^{r}_{c,d}, p^{n}_{c,d}, p^{s}_{c,d})
\end{equation}

\noindent \textit{\textbf{+Verification} (Ideal Reranker Model)} combines semantic relevance and verification feedback to refine evidential retrieval. The final retrieval score is calculated by summing the semantic score ($s^{r}_{c,d}$) and the verification probabilities ($p^{r}_{c,d}$ and $p^{n}_{c,d}$), followed by a normalisation step to ensure the score remains within (0,1), formulated as:
\begin{equation}
\label{equa:+v_euqation}
s^{r+v}_{c,d} = 1/2 * (s^{r}_{c,d} + p^{r}_{c,d} + p^{s}_{c,d}) \in (0,1)
\end{equation}
This formulation ensures that documents contributing to support or refute labels receive higher retrieval scores, enhancing the evidential quality of retrieved documents.

To evaluate retrieval effectiveness, we use \textit{Recall@k} (R@k), a common evaluation approach that measures the proportion of relevant evidence successfully retrieved within the top $k$ results.

\paragraph{\textbf{Datasets}}
We conduct our evaluation using datasets including:
\textbf{(1) SciFact}  \cite{wadden-etal-2020-fact}: A corpus of 5,183 abstracts from scientific articles, with 809/300/300 samples for train, validation and test sets. The test set is not publicly accessible.
\textbf{(2) SciFact-Open}  \cite{wadden-etal-2022-scifact}: An extended version of SciFact with 500K abstracts, re-annotating evidence documents for 279 claims from the original SciFact test set. 
\textbf{(3) Check-COVID}  \cite{wang-etal-2023-check-covid}: A COVID-19-specific fact-checking dataset, containing 347 abstracts from CORD-19 journal articles and 1,504 expert annotated news-related claims.

MultiVerS is trained on the SciFact train set to create a verifier model that provides verification feedback. Since the SciFact test set is inaccessible, we evaluate document evidence retrieval on SciFact-Open and full Check-COVID.

\paragraph{\textbf{Results}}

\begin{table}[t]
\caption{Retrieval result on SciFact-Open and Check-COVID}
\vspace{-10pt}
\begin{minipage}{0.46\textwidth}
\resizebox{\textwidth}{!}{

\begin{tabular}{ccccccc}
\hline
SciFact-Open  & R@50                                                                                    & R@20                                                                                    & R@10                                                                                    & R@5                                                                                     & R@3                                                                                     & R@1                                                                                     \\ \hline
BM25          & 66.09                                                                                   & 54.78                                                                                  & 45.22                                                                                  & 38.04                                                                                  & 30.87                                                                                  & 20.22                                                                                    \\
monoT5-3B     & 88.91                                                                                   & 79.13                                                                                   & 71.09                                                                                   & 57.17                                                                                   & 48.26                                                                                   & 31.09                                                                                   \\ 
+Verification     & \textbf{91.96}                                                                          & \textbf{81.95}                                                                          & \textbf{71.30}                                                                          & \textbf{62.61}                                                                          & \textbf{52.83}                                                                          & \textbf{32.17}                                                                                 \\ \hline
Check-COVID   & R@50                                                                                    & R@20                                                                                    & R@10                                                                                    & R@5                                                                                     & R@3                                                                                     & R@1                                                                                     \\ \hline
BM25          & 87.91                                                                                   & 81.96                                                                                   & 75.02                                                                                   & 67.59                                                                                   & 61.35                                                                                   & 46.18                                                                                   \\
monoT5-3B     & 95.84                                                                                   & 93.16                                                                                   & 89.49                                                                                   & 82.06                                                                                   & 74.93                                                                                   & 58.28                                                                                \\
+Verification        & \textbf{96.13}       & \textbf{94.55} & \textbf{91.48} & \textbf{84.04} & \textbf{77.80} & \textbf{61.84}

\\ \hline
\end{tabular}
}
\end{minipage}
\label{tab:result_recall}
\vspace{-5pt}
\end{table}

The integration of verification feedback consistently enhanced document evidence retrieval, as +Verification outperformed monoT5-3B across nearly all cut-off thresholds in both SciFact-Open and Check-COVID (Table~\ref{tab:result_recall}). The improvements are particularly evident at lower cut-offs, where retrieving the most relevant evidence is crucial. In SciFact-Open, Recall@5 and Recall@3 increased from 57.17\% to 62.61\% and from 48.26\% to 52.83\%, respectively. Similarly, in Check-COVID, these metrics improved from 82.06\% to 84.04\% and from 74.93\% to 77.80\%. These improvements are particularly meaningful given that the average number of gold evidence documents per claim is 1 in Check-COVID and approximately 2 in SciFact-Open. To illustrate this improvement, we conducted a case study examining how different retrieval methods ranked specific evidence documents (Table~\ref{tab:case_study}). Compared to monoT5-3B, +Verification successfully elevated the ranks of E2, E3 and E5, retrieving three additional pieces of evidence within the top 5 results. Notably, E5, ranked 302nd by monoT5-3B, was effectively rescued by adding verification feedback in +Verification, demonstrating the value of integrating verification-informed retrieval signals.

\begin{table}[t]
\caption{\small Evidence positions in the retrieved list. We select an example from the SciFact-Open dataset.
Claim: Female carriers of the Apolipoprotein E4 (APOE4) allele have a reduced risk for Alzheimer's disease.\space Gold Evidence: [E1,E2,E3,E4,E5] 
}
\vspace{-10pt}
\begin{tabular}{llllll}
\hline
reranker model  & E1  & E2   & E3   & E4  & \textbf{E5}    \\ \hline

BM25  & 838th & 141th & 7th  & 163th & \textbf{67th}   \\ 
monot5-3B  & 1st & 8th & 16th  & 2nd & \textbf{302nd}   \\ 
+Verification & 1st & 3rd & 5th  & 2nd & \textbf{27th}  \\\hline
\end{tabular}
\label{tab:case_study}
\vspace{-10pt}
\end{table}

These findings highlight a promising research direction: shifting from semantic-only retrieval towards evidence-aware retrieval, where retrieval models explicitly account for evidential value. Our results suggest that leveraging well-performing verification models can help refine retrieval systems by distinguishing between purely semantic relevance and plausible evidential relevance among unannotated documents. Furthermore, to continually improve evidence-aware retrieval, we propose the development of tailored IR systems capable of identifying evidential information, thereby enhancing evidence retrieval for scientific fact-checking.

\subsection*{\textit{\normalsize{RD.1. Benchmark tailored IR system for fact-checking}}}
The preliminary study presented in this work outlined a framework to enhance evidence retrieval beyond only semantic relevance. To overcome the limitations of existing IR systems in scientific fact-checking scenarios, it is imperative to develop specialised IR systems capable of handling the specific challenges of verification tasks. However, training an IR system on a single fact-checking dataset risks poor generalizability and potential overfitting, particularly due to data imbalance, a common issue in the relatively small datasets characteristic of scientific fact-checking  \cite{zeng2021automated,vladika-matthes-2023-scientific}. Furthermore, poor verification performance deteriorates retrieval accuracy, creating a vicious feedback loop that further degrades overall system effectiveness. A multi-pronged strategy could mitigate these challenges by \textit{pooling verification signals} from various high-performing verifier models, \textit{leveraging large-scale datasets} such as FEVER  \cite{thorne-etal-2018-fever} to improve training robustness, and \textit{providing a shared retrieval checkpoint} enable subsequent studies to fine-tune the model for specific scenarios or datasets while reducing training cost. 
Recent work  \cite{li2024corpuslm,salemi2024towards} has explored unified retrieval models for knowledge-intensive NLP tasks, focusing on retrieval quality and downstream task utility  \cite{salemi2024evaluating,zamani2024stochastic}, including question answering (QA) and fact-checking. Similarly, we propose a verification-driven IR system for evidence retrieval, which explicitly incorporates evidential informatics. This approach follows a two-step training paradigm: general pre-training on large, diverse datasets followed by domain-specific fine-tuning. This approach balances scalability and domain specificity, ensuring IR models are both robust across different contexts and highly effective in targeted fact-checking applications.
Additionally, a corresponding benchmark should employ a diverse set of evaluation metrics beyond for the fact-checking task to ensure comprehensive assessment of performance within fact-checking \cite{akhtar2024ev2r}. These metrics could include \textit{verification accuracy}, reflecting the downstream utility of retrieved evidence; \textit{decision latency}, measuring the computational efficiency of retrieval models; and \textit{robustness to real-world conditions} such as noisy data and incomplete evidence, to improve system resilience.

Integrating verification feedback into evidence retrieval improves relevance assessment beyond binary labels, enhancing retrieval performance. Future research should focus on developing a benchmark IR system tailored for fact-checking, incorporating fine-grained relevance labels and verification-driven retrieval models. A scalable pre-training and fine-tuning approach has the potential to improve retrieval robustness and generalizability thereby producing more accurate and efficient fact-checking systems.

\section{Imbalanced resources of scientific topics}
\label{sec:imbalance_resources}
In existing general fact-checking datasets, such as FEVER which is based on Wikipedia, the distribution of gold evidence per claim is relatively even and sufficient. However, a significant imbalance of evidence is observed in scientific fact-checking.
While the SciFact corpus (\textasciitilde5K documents) maintains a relatively balanced number of evidence documents per claim, this balance was disrupted when the dataset was expanded to create SciFact-Open (\textasciitilde500K documents). In this larger corpus, the majority of claims have none or only one piece of supporting evidence while others have over.
Claims related to less-researched topics are generally associated with fewer scientific publications, as illustrated by the examples in Table~\ref{tab:imbalance_evidence}.

\begin{table}[t]
\caption{\small Sufficient-evidence claim and none-evidence claim examples in SciFact-Open. `Evidence' is the number of evidence in SciFact-Open corpus and `Entities' is the number of entities by searching bold-keyword in PubMed.}
\vspace{-5pt}
\begin{tabular}{lcc}
\hline
\textbf{Claim}                                                                               & \multicolumn{1}{l}{\textbf{Evidence}} & \multicolumn{1}{l}{\textbf{Entities}} \\ \hline
\begin{tabular}[c]{@{}l@{}}\textbf{Obesity} is determined in part by \\ genetic factors.\end{tabular} & 24                                    & 499k                                            \\ \hline
\begin{tabular}[c]{@{}l@{}}\textbf{LRBA} controls CTLA - 4 expression.\end{tabular}                 & 0                                     & 0.27k                                             \\ \hline
\end{tabular}

\vspace{-5pt}

\label{tab:imbalance_evidence}
\end{table}

However, most fact-checking systems do not explicitly account for evidence imbalance. A common approach is to use a fixed retrieval cut-off (i.e., selecting a predefined number of top-ranked documents for verification). One of the most inefficient approaches is setting the cut-off equal to the maximum number of evidence per claim in the dataset, ensuring that all possible evidence is retrieved.
This heuristic has not previously caused major issues since general fact-checking datasets contain a relatively balanced number of supporting documents per claim. However, the imbalance in SciFact-Open suggests that the simple approach may not be suitable for open-domain scientific fact-checking with two major drawbacks:

(1) \textbf{Inefficiency.} Although the maximum number of gold evidence documents in the SciFact-Open dataset is 24, less than one third of claims have more than two. Using the maximum number as a cut-off would be inefficient due to the large number of documents that would have to be processed by the verifier. 

(2) \textbf{Inaccuracy.} Introducing irrelevant evidence into downstream verification degrades fact-checking performance, whether the noise is semantically related or completely random  \cite{sauchuk2022role} (as discussed in Section~\ref{sec:beyond_semantics}).

To address these challenges in the current and future studies of scientific fact-checking, we proposed a research direction based on flexible cut-off strategy for retrieving evidence based on claim characteristics.

\subsection*{\textit{\normalsize{RD.2. Flexible cut-off for Retrieved Evidence}}}

Ranked List Truncation (RLT) refers to the task of selecting an optimal prefix of a ranked list of retrieved documents, with the goal of balancing retrieval effectiveness and efficiency. Prior work explores both heuristic and learned approaches, using either relevance labels or features derived from score distributions to determine the cut-off point~\cite{meng2024ranked,ma2022incorporating,wang2022mtcut,wu2021learning,bahri2020choppy,lien2019assumption}.
A related line of work is stopping methods in technology-assisted review (TAR) \cite{bin2024rlstop,hezam-stevenson-2023-combining,stevenson2023stopping}, which aim to retrieve as much relevant information as possible while minimising the effort spent on examining irrelevant documents.
Both approaches aim to optimise an expected metric over candidate cut positions, typically using metrics such as $F1@k$ or $recall@k$.
The datasets used in these prior studies on RLT and stopping methods, such as CLEF and TREC \cite{Kanoulas2018CLEF2T,kanoulas2018clef,kanoulas2019clef,cormack2014evaluation,grossman2016trec,craswell2020overview,craswell2021overviewtrec2020deep}, exhibit imbalance but typically contain enough relevant items per query to support recall-based supervision and evaluation.
This dependence on sufficient relevance labels becomes problematic in fact-checking scenarios, where gold evidence documents are typically rare, making both recall-based stopping and supervised RLT approaches unsuited to this problem.

To explore whether relevance score distributions indicate evidence sufficiency, we compare well-studied and less-studied claims from SciFact-Open. Following  Wadden et. al. \cite{wadden-etal-2022-scifact}, claims with four or more gold evidence documents are considered to be well-studied and those with none to be less-studied. Using monoT5-3B, we compute several statistics over the ranked document scores, including first-document relevance, average and total scores, score decay, and the initial-to-final score ratio.
Table~\ref{tab:statistic_analysis} shows that well-studied claims tend to have higher top-ranked scores and sharper decay patterns, suggesting that relevance distributions may serve as indicators of evidence sufficiency.

\begin{table}[h]
\caption{Statistical analysis of average relevance scores for well-studied and less-studied claims. `I/F' ratio is Initial-to-Final ratio. `Exp k' denotes the exponential decay factor k.}
\begin{tabular}{cccccc}
\hline
\textbf{Metric} & \textbf{1st Doc} & \textbf{Mean} & \textbf{Sum} & \textbf{I/F ratio} & \textbf{Exp k} \\ \hline
\textbf{Less-}  & 0.941            & 0.502         & 25.097       & 3.514              & 1.544          \\
\textbf{Well-}  & 0.995            & 0.741         & 37.052       & 1.963              & 0.651          \\ \hline
\end{tabular}

\label{tab:statistic_analysis}
\end{table}

Based on these findings, one possible direction is to leverage existing techniques to estimate whether a claim is less-studied or well-studied. A prediction module could utilise statistical features of relevance distribution such as those presented in Table~\ref{tab:statistic_analysis}. In addition, metadata such as retrieved entity counts in PubMed (Table~\ref{tab:imbalance_evidence}) can serve as auxiliary signals to refine the prediction. Claims predicted as less-studied -- e.g., with low total relevance or steep score decay -- may be assigned smaller cut-offs to reduce verification cost, while RLT and stopping techniques could be applied to well-studied claims where concentrated high scores suggest richer evidence.

While this naive strategy relies on heuristic features, it does not explicitly optimise verification performance. To address this, future approaches could explore learning a cut-off policy using feedback from the verification stage. Specifically, truncation points may be selected based on reward signals, such as whether the claim is correctly verified or the confidence of the verifier. This would bypass the need for relevance-labelled supervision, which is often infeasible in scientific fact-checking due to sparse annotations. Inspired by prior RLT and stopping method work, such a learned policy could optimise both efficiency and factual accuracy by aligning truncation decisions with downstream verification performance.

\section{Time and Citation}
\label{sec:time_citation}
General fact-checking evidence corpus such as Wikipedia and fact-checking websites, often lack sentence-level evidence timestamps, making it difficult to determine the original publish time of sentence evidence in verification and hindering the development of time-aware retrieval methods. 
Timeliness is important in fact-checking, but in science, evolving evidence makes outdated studies particularly problematic. For instance, early COVID-19 treatment studies were later refuted, and outdated evidence may lead to harmful decisions.
This section discusses whether scientific publications are more suitable for time-aware fact-checking and explores possible ways to leverage their inherent temporal characteristics.

\subsection{Timeliness of evidence}
Unlike general fact-checking, where historical and static facts remain unchanged, scientific knowledge continuously evolves. This fundamental difference necessitates time-aware retrieval in scientific fact-checking to ensure that retrieved evidence remain valid and reflective of the latest scientific consensus. A time-sensitive retrieval mechanism should prioritise recent publications to ensure that fact-checking systems incorporate the most up-to-date methodologies and factual updates.  
This is especially crucial in fields like healthcare, where relying on outdated information could lead to misleading conclusions or incorrect decisions. For example, during a rapidly evolving pandemic, a medical treatment initially considered effective might later be deemed unreliable. This section discusses the challenges and opportunities of integrating the evidence timestamp into scientific fact-checking.

\begin{table}[t]
\caption{Results of health QA task considering the different thresholds of the published time of literature  \cite{vladika-matthes-2024-improving}} 
\begin{tabular}{cccc}
\hline
\textbf{Year}      & \textbf{Precision} & \textbf{Recall} & \textbf{F1 score} \\ \hline
$\ge$\textbf{2020} & 59.7               & \textbf{60.3}            & \textbf{58.7}              \\
$\ge$\textbf{2018} & 59.6               & 58.0            & 57.9              \\
$\ge$\textbf{2015} & 61.1               & 56.0            & 53.9              \\
$\ge$\textbf{2010} & 63.4               & 55.6            & 52.8              \\
$\ge$\textbf{2005} & \textbf{68.1}               & 56.5            & 52.0              \\
$\ge$\textbf{2000} & 66.1               & 56.8            & 51.8              \\
$\ge$\textbf{1990} & 65.6               & 55.4            & 51.3              \\
$\ge$\textbf{1980} & 64.2               & 54.7            & 50.0              \\ \hline
\end{tabular}
\label{tab:time-qa}
\end{table}

Scientific fact-checking aims to find evidential information in the literature to verify a claim. Intuitively, considering outdated literature negatively affects verification. Research in healthcare question answering (QA) has demonstrated that time-aware retrieval improves system performance  \cite{vladika-matthes-2024-improving}, as shown in Table~\ref{tab:time-qa}.  
The F1 score of the healthcare QA system improves as the publication year of evidence documents becomes more recent. By extension, scientific fact-checking on a large-scale corpus may also suffer from incorporating outdated evidence. These findings highlight the need for time-aware filtering in scientific fact-checking systems to enhance reliability.
Fact-checking datasets could explicitly incorporate timestamps as metadata to facilitate research into temporal relevance in retrieval and verification. A general fact-checking study  \cite{augenstein-etal-2019-multifc} collected the `timestamp of last update' of claims and evidential documents, allowing later research  \cite{allein-etal-2023-implicit} to explore the impact of temporal data on verification. This study found that a time-aware system achieved a 15\% improvement in macro F1 score, underscoring the importance of temporal information. However, this work was limited by: (1) \textit{focusing on verification only}, where evidence documents had already been retrieved in a separate initial step, without ensuring that retrieval prioritised high-quality and temporally relevant evidence, and (2) \textit{the use of fragmented evidence} -- the general fact-checking datasets usually only contain small snippets of web documents, omitting many important time expressions present in full texts. 

Given these findings, it is essential to develop a fact-checking system that explicitly incorporates temporal awareness across both retrieval and verification stages to meet real-world applications.

\subsection*{\textit{\normalsize{{RD.3. Time-Aware Retrieval and Verification}}}}
Traditional fact-checking approaches often handle conflicting evidence for a single claim by assigning neutral veracity labels, such as ``mixture," ``unproven," or ``not enough information"  \cite{guo2022survey,vladika-matthes-2023-scientific}. However, conflicting evidence often arises due to outdated studies included in the retrieval process, which introduces noise and adversely affects prediction accuracy \cite{vladika-matthes-2024-improving}. This issue has been observed in healthcare QA systems, where outdated evidence degrades performance  \cite{vladika-matthes-2024-improving}. To address this issue, we propose a time-aware approach that incorporates temporal information into both retrieval and verification stages:

\textbf{1.Retrieval Stage}: Outdated evidence should be filtered or deprioritised during retrieval to ensure that the retrieved evidence set is temporally aligned with the latest scientific findings.

\textbf{2.Verification Stage}: After filtering by time-aware retrieval, all retrieved evidence should be considered but with differentiated weighting based on temporal relevance. Recent evidence should be prioritised through higher weights, while older evidence should serve as supplementary or contextual information rather than primary evidence.

By processing outdated evidence differently across retrieval and verification components, this direction explores how temporal information can reduce noise and improve the reliability of scientific fact-checking systems.

\subsection{Indirect evidence through citation}
\label{subsec:timestamp}
General fact-checking faces a number of challenges when attempting to determine the timeliness of evidence: 
(1) \textit{Lack of publication timestamp}  \cite{allein-etal-2023-implicit}. Many fact-checking sources, such as Wikipedia and fact-checking websites, do not provide precise publication dates for individual sentences or paragraphs. Instead, they record only the last edited timestamp, which does not accurately reflect when a fact was first published. 
(2) \textit{Tracking the origin of evidence}. Evidence is often copied or paraphrased across multiple sources, making it difficult to determine the original publication date of a statement. 
(3) \textit{Search engine bias in retrieval}. Pre-established fact-checking datasets retrieve evidence using top-k search engine results, where ranking mechanisms may prioritise recent documents due to time-aware ranking biases. This can misrepresent the actual chronology of claims and lead to fragmented evidence, making time extraction unreliable.

While these challenges also affect scientific fact-checking, they can be naturally mitigated by leveraging full-text papers rather than abstract-only sources. The main advantages of using full-text papers include: (1) \textit{Explicit publication metadata}: Each piece of literature has a clearly defined publication date, ensuring accurate temporal tracking. (2) \textit{Citation tracing for indirect evidence}: Mandatory citation rules in academic publications facilitate source tracing, even when statements are referenced indirectly. (3) \textit{Structured nature of academic papers}: provides an indication of the origin of statements. For example, evidence in the background section typically references prior studies, whereas those in the abstract or results sections represent findings from the current publication. Given these inherent advantages, incorporating timestamps as metadata offers a promising research direction for full-paper-based scientific fact-checking. Exploring the temporal dynamics of claims and evidence should further enhance retrieval accuracy. Additionally, citation tracking to trace the original source of paraphrased evidence, ensuring the first-published timestamp is accurately recorded. 

Citation-based tracking provides a promising approach to estimating evidence timestamps. However, the widespread presence of multiple citations in scientific literature makes it difficult to identify which references should be tracked, increasing computational costs and reducing efficiency. Moreover, indirect citations and paraphrased references, particularly in introductory sections, further obscure the retrieval of the first-published source. To address these issues, we propose the following research direction to develop a more effective approach for citation tracking and timestamp attribution.

\subsection*{\textit{\normalsize{{RD.4. Citation-Based Evidence Tracking}}}}

\noindent Intuitively, self-contained evidence refers to information directly presented in the body of the current paper, while cited evidence is derived from external sources referenced by the paper. 
To explore this distinction, we analysed 22 accessible full papers out of 24 gold evidence for the claim in Table~\ref{tab:case_study}. We prompted GPT-4o to search for supporting/refuting evidence and determine whether each piece of evidence was paraphrased/summarised from a citation or is self-contained.

\begin{figure}[ht]
  \centering
  \includegraphics[width=0.8\linewidth]{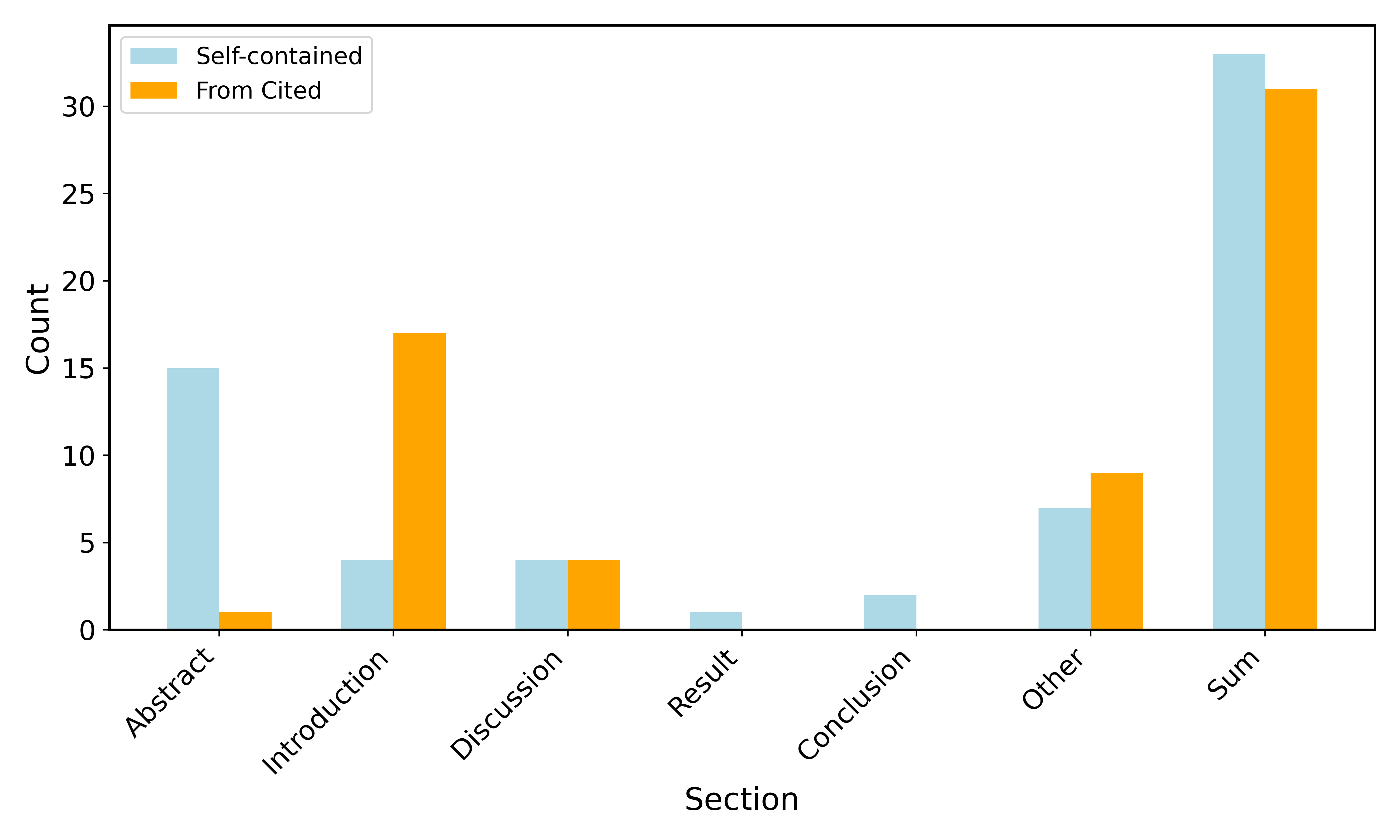}
  \vspace{-10pt}
  \caption{Evidence sources in scientific literature.}
  \Description{Analysis of two different evidence sources for a single claim.}
\label{fig:cited_evidence}
\end{figure}

The result shown in Figure~\ref{fig:cited_evidence} reveals, as expected, that cited evidence is predominantly located in the introduction, while self-contained evidence is more common in the results and conclusion sections. 
While we observed a few inaccurate outcomes (e.g., one `From Cited' evidence appearing in `Abstract' is misjudged), the overall distribution remains discernible and interpretable.
In addition, we also observed that sentences in scientific literature frequently contain multiple citations, making it costly to manually extract the original timestamp of cited evidence. To address this challenge, we propose a two-step approach:

\textbf{1. Identify track-worthy citations.} 
Not all citations are equally important for fact verification. Track-worthy citations should include: \textit{conclusive evidence} directly influences claim verification and \textit{plausible evidence} that may impact claim assessment  \cite{wright-augenstein-2021-citeworth}. 
Since checking every cited document is expensive, an initial filtering step is required. A potential solution is to rank citations based on their relevance and function. 
Citation recommendation  \cite{farber2020citation,gu2022local} is a similar task to identify relevant publications for a given statement using retrieval models. Beyond relevance, the function of citation can be referred to as a signal to adjust priority. Citations are classified into eight categories: background, motivation, uses, extends, similarities, differences, compare/contrast, and future work  \cite{ikoma-matsubara-2023-use}. While single evidence has multiple citations, the function of citation can help identify the citations that align with the evidence. For example, for predicted evidence `Experiments show model A outperforms previous SOTA model B [citation 1,2,3]', cited papers for `model A/B' in the `introduction' function are less check-worthy than the citation in the `compare/contrast' function. By prioritising high-impact citations, retrieval costs can be significantly reduced.

\textbf{2. Track the original timestamp of evidence.}   
Once track-worthy citations are identified, the next step is to trace the original timestamp of cited evidence via direct and indirect citation tracking. \textit{Direct citation tracking} can be applied if the publication date of the cited paper is straightforward to retrieve. However, some evidence is paraphrased or indirectly cited, requiring a deep tracking mechanism to trace the citation path. 
Citation graph analysis  \cite{viswanathan-etal-2021-citationie,buneman2021data} can help map citation paths using directed graphs and applying search algorithms to identify the earliest relevant source. A recent study  \cite{zhang2024detecting} found that reference errors -- references do not include information to support statement -- frequently appear ranging from 11\% to 41\% across domains. Addressing these errors introduces a sub-task for scientific fact-checking: verifying whether a cited reference truly supports the claim. To ensure feasibility, an early explorative study can assume that scientific literature generally adheres to citation conventions, preventing infinite citation loops in verification.

In summary, to explore time-aware fact-checking for scientific literature, we propose two potential research directions: (1) Integrating temporal information into retrieval and verification to handle outdated evidence and (2) Developing citation-based tracking methods to identify the original source and timestamp of evidence.
These directions provide a basis for studying the impact of time-aware mechanisms in scientific fact-checking systems.

\section{Structured Long-Context Evidence Retrieval}
\label{sec:structure}

Existing scientific fact-checking datasets construct evidence corpora using fragmented sentences, paragraphs, or abstracts \cite{wadden-etal-2020-fact,mohr-etal-2022-covert,wang-etal-2023-check-covid,bulian2020climate,saakyan-etal-2021-covid,sarrouti-etal-2021-evidence-based,kotonya-toni-2020-explainable-automated}. However, scientific literature is typically presented in structured, visually rich formats, often as PDF documents, where different sections serve distinct functions: abstracts, results, and conclusions summarise key findings, while background and introduction sections provide prior research context. With the diverse and unique functionalities of scientific literature components, this section explores challenges and potential research directions for advancing scientific fact-checking at the full-paper level. 
Existing fact-checking pipelines typically follow a document retrieval and then sentence selection paradigm  \cite{guo2022survey,zeng2021automated,vladika-matthes-2023-scientific}. For general fact-checking, evidence retrieval often uses top-ranked sentences from top-ranked documents, treating them as self-contained evidence units. However, this approach neglects long-range context, as using the extracted sentences ignores surrounding information to support verification. 
In contrast, scientific documents exhibit higher document-level consistency of verdict, commonly one paper having a sole standpoint to a given question, making document-level processing necessary for scientific fact-checking \cite{li2021paragraph,zhang-etal-2021-abstract,wadden-etal-2022-multivers}. 

\begin{figure}[t]
  \centering
  \includegraphics[width=0.9\linewidth]{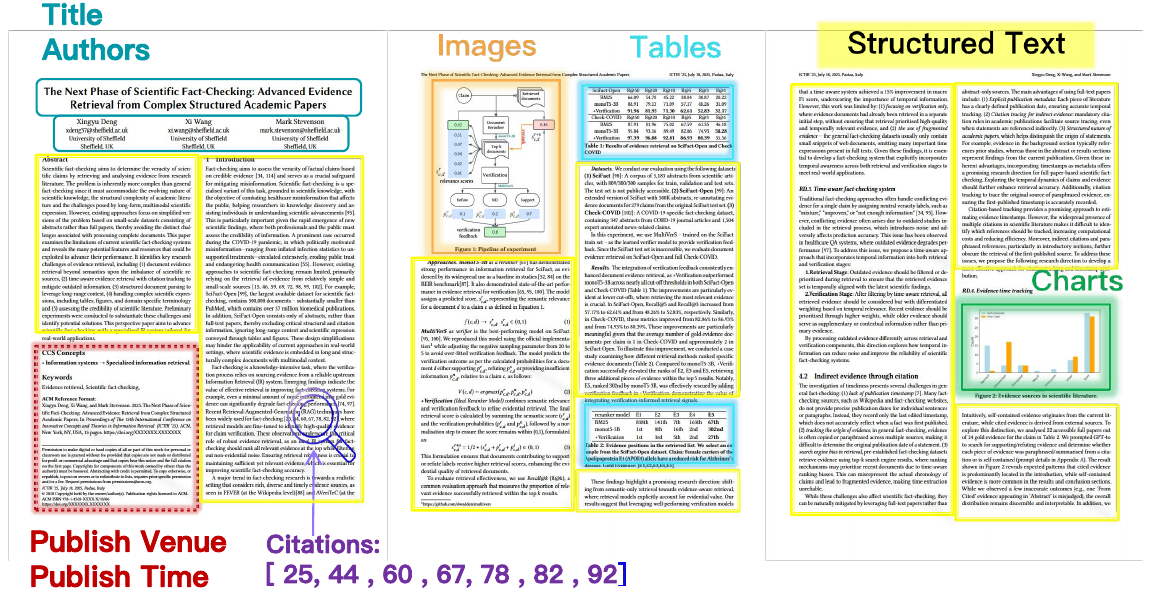}
  \caption{An example for parsing the scientific literature}
  \Description{An example to parse scientific literature.}
\label{fig:demo_parser}
\vspace{-10pt}
\end{figure}

LLMs have recently demonstrated growing capability to process long contexts. However, it remains challenging to fact-check an entire full-text document in a single pass  \cite{wang-etal-2024-factcheck}. LLMs are prone to hallucinating content that is not grounded in the provided documents \cite{tang-etal-2024-minicheck,chiesurin-etal-2023-dangers,chen2024benchmarking,adlakha-etal-2024-evaluating} and are often susceptible to distraction from irrelevant context \cite{shi2023large}. Their reasoning capabilities also degrade as text length increases \cite{karpinska-etal-2024-one}. A potential solution is context distillation, as used in retrieval-augmented generation (RAG), to filter high-quality context and mitigate hallucination, which improves the performance of downstream QA tasks \cite{wang2023learning}. However, unlike the QA task, fact-checking requires explicit retrieval of explicit. Filtering context may remove crucial supporting evidence, leading to incomplete verification. Moreover, much of the historical scientific literature exists in PDF format. Although recent multimodal LLMs are capable of consuming PDFs and conducting reasoning tasks, their capabilities are still limited to surface-level understanding. For example, they frequently fail to capture cross-page content and complex layout structures~\cite{van2023document,ma2024mmlongbenchdocbenchmarkinglongcontextdocument,tanaka2023slidevqa}. By prompting GPT-4o to locate evidence in PDF literature with the results presented in Section~\ref{subsec:timestamp}, we observed that it mislocated evidence in incorrect sections or non-existent sections. These limitations indicate that existing LLMs are not capable of supporting end-to-end fact-checking for long-context scientific literature -- not only for raw PDFs but also for plain-text documents.

In addition to their length, scientific papers follow a structured format that introduces additional challenges for verification. Background and introduction sections often cite prior work, which may conflict with conclusions drawn later, while discussion sections highlight limitations that can cast doubt on earlier findings. 
These internally inconsistent signals may introduce misleading information, demanding reasoning that accounts for both the factual assertions and the functional roles of different sections.
Parsing scientific documents into structured units enables a modular pipeline where retrieval and verification can be independently optimised. 
This modular structure facilitates interpretability, robustness, and denoising of conflicting or irrelevant content.
Layout-aware tools offer a practical foundation for such structure-aware processing~\cite{chen-etal-2023-layout,fok2023scim,10.1145/3659096,huang-etal-2022-lightweight,shen-etal-2022-vila}, as illustrated in Figure~\ref{fig:demo_parser}.

To address these challenges, we advocate for a retrieval framework that explicitly considers the document structure commonly found within scientific literature. Such a system should identify targeted evidence, capture long-range context across sections, and suppress irrelevant or conflicting content. We next outline a direction toward adaptive, section-aware retrieval strategies designed to meet these requirements.

\subsection*{\textit{\normalsize{{RD.5. Adaptive Section-Aware Evidence Retrieval}}}}
Scientific literature follows a structured format where different sections serve distinct functions, presenting challenges for traditional evidence retrieval and verification in fact-checking systems. Existing retrieval methods often operate at the sentence or paragraph level, neglecting the long-range context and the structured nature of scientific documents. Additionally, large language models (LLMs) struggle to accurately associate claims with the appropriate sections, leading to potential misinterpretations and inconsistencies.

A promising research direction is \textbf{adaptive section-aware evidence retrieval}, which dynamically adjusts retrieval strategies based on document structure and claim types. This approach consists of two key components:

\textbf{Claim-evidence matching}. Claims should first be matched to the most relevant sections. For instance, experiment-driven claims, such as “X method improves accuracy compared to Y,” should primarily retrieve evidence from the `Results' and `Conclusion' sections, as these contain empirical findings. In contrast, background or theoretical claims, such as “X method is widely used in Y applications,” should focus on the Introduction and Background sections, which provide foundational knowledge. Prioritising section-aware evidence retrieval helps filter irrelevant context and reduces retrieval noise.

\textbf{Contextual expansion}. After retrieving primary evidence, the system should augment it with relevant contextual information from other sections to improve interpretability. For example, methodological details from the `Analysis' section can provide additional support for experimental claims, while historical context from the Background section can clarify theoretical claims. Access to the full document allows for retrieving finer-grained evidence or complementary details that fragmented approaches may overlook.  

In summary, an effective adaptive section-aware retrieval system must overcome challenges in accurately parsing document structures, prioritising relevant sections based on claim types, and efficiently handling conflicting evidence. By integrating structured document parsing, hierarchical retrieval strategies, and context-aware reasoning, future systems can leverage richer evidence from full-paper scientific literature while reducing LLM hallucinations. Advancing these techniques will enhance the reliability and interpretability of scientific fact-checking systems.

\section{Multimodal content in Science}

Scientific literature often conveys key evidence using non-textual elements such as tables, charts, and figures, which are commonly used to present experimental results, statistical analyses, and theoretical models, as shown in Figure~\ref{fig:demo_parser}. For full-text scientific fact-checking, especially across various fields of science and technology, it is crucial to move beyond text and accurately interpret these structured elements to ensure comprehensive verification.
Fact-checking and misinformation detection on individual modalities -- such as figures~\cite{tonglet-etal-2024-image,papadopoulos2024verite,papadopoulos2023synthetic,qi2024sniffer,abdelnabi2022open,yao2023end}, charts~\cite{akhtar-etal-2023-reading,akhtar-etal-2024-chartcheck}, and tables~\cite{bhagavatula2013methods,schlichtkrull-etal-2021-joint,dong2024large,2019TabFactA,shi-etal-2021-logic,akhtar-etal-2022-pubhealthtab,gu-etal-2022-pasta,ye2023large,lu-etal-2023-scitab} -- has been studied independently~\cite{akhtar-etal-2023-multimodal}. To improve tabular reasoning, transformer-based approaches such as TAPAS~\cite{herzig-etal-2020-tapas} and Table-BERT~\cite{yin-etal-2020-tabert} have been developed. However, these techniques perform poorly in SCITAB~\cite{lu-etal-2023-scitab}, a dataset for scientific fact-checking on tables, with results barely above random. One possible cause is the lack of contextual grounding for tables, which are rarely self-contained.
Similarly, figures and charts in scientific papers often require surrounding textual explanations for correct interpretation. Recent datasets like AVerImaTeC \cite{cao2025averimatecdatasetautomaticverification} address image-text verification using web-sourced data, but scientific domains present greater challenges: figures are densely structured, often span multiple sections, and require domain-specific understanding capability.

We argue that scientific fact-checking should shift toward full-paper analysis, where structured elements are interpreted alongside their textual context. Unlike standalone multimodal models, document-level processing enables cross-referencing between figures/tables/charts and their descriptions, facilitating more faithful and complete verification.

\subsection*{\textit{\normalsize{{RD.6. Multi-modal Evidence Alignment}}}}

Scientific fact-checking requires integrating evidence across multiple modalities, including text, tables, and figures, to ensure consistency and completeness. Recent multimodal information retrieval datasets in the scientific domain \cite{roberts2024scifibench,wu-etal-2024-scimmir} provide aligned pairs of textual and structured content, offering a foundation for cross-modal reasoning. However, in real scientific documents, structured elements, such as figures and tables, are not always located close to their descriptive text, making alignment a non-trivial challenge.

A promising direction is to explicitly align structured elements with their corresponding textual explanations within the same document. In scientific articles, tables and figures are typically explained through captions or surrounding sentences. Layout-aware parsing techniques can help identify these elements and link them to relevant text spans. Once aligned, their contents can be jointly encoded, enabling claim verification that draws on both structured data and contextual text. This alignment facilitates more coherent retrieval and reasoning across modalities, improving the reliability of multimodal fact-checking.

This unified framework contrasts with traditional multimodal systems that process each modality in isolation. Effective alignment demands progress in scientific document understanding~\cite{ma2024mmlongbenchdocbenchmarkinglongcontextdocument,cho2024m3docrag}, visual structure parsing~\cite{fok2023scim}, and domain-specific retrieval~\cite{roberts2024scifibench}. Integrating these efforts will support full-document, multimodal verification, where structured evidence is faithfully grounded in its textual context.

\section{Credibility of scientific literature}
Existing evidence retrieval models often favour high-ranking documents based on semantic relevance, often overlooking scientific rigour  \cite{vladika-matthes-2024-comparing}. As a result, low-quality documents may be retrieved as evidence, undermining the credibility of scientific fact-checking. The reliability of a claim verification process is inherently tied to the quality of the supporting literature, making evidence credibility a crucial factor in scientific fact-checking. 

In the domain of scientific literature, credibility assessment is influenced by multiple factors, including \textit{peer-review status}, which ensures methodological scrutiny, \textit{citation impact} which indicates how influential a study is within its field, and \textit{experimental rigour}, reflecting the robustness of a study's methodology. 
However, the proliferation of non-peer-reviewed manuscripts and publications from venues with varying editorial standards, particularly in open-access repositories, poses a growing challenge. Such sources may lack the rigorous methodological scrutiny necessary to ensure reliable scientific conclusions.
Therefore, incorporating additional quality indicators is essential for enhancing the robustness of scientific fact-checking. 

\subsection*{\textit{\normalsize{{RD.7. Extending Indicators of Evidence Quality}}}}
Evaluating scientific literature quality extends beyond content reliability, and should consider factors including venue reputation, methodological rigour, and experimental transparency. 
High-impact journals and prestigious conferences generally enforce stringent peer-review standards, contributing to the credibility of published research. 
Similarly, the expertise and prior contributions of an author, particularly in reputable venues, can provide further insight into the credibility of a study.
In addition to traditional metadata, emerging indicators such as replication status, data availability, and adherence to reporting guidelines can further reflect methodological soundness.
Although metadata-based credibility assessment is a useful heuristic, it is not foolproof. 
For example, selective reporting and statistical manipulation still exist, as some widely cited studies have later been retracted due to methodological flaws \cite{fletcher2025predicting}. 
While the integration of such metadata remains a reasonable approach, as these indicators generally correlate with literature quality, their limitations must be acknowledged, given that even widely cited studies can occasionally be subject to retraction due to undetected methodological flaws.

\section{Scientific terminology complexity}
\label{sec:terminology_complex}
Scientific fact-checking systems often encounter challenges when aligning claims with supporting evidence due to mismatches in terminology granularity  \cite{wadden-etal-2022-scifact,west2021misinformation}. The prevalence of hierarchical and synonymous scientific terms introduces significant challenges in fact verification. Many concepts exist at multiple levels of specificity, where broader categories encompass more specific subtypes, leading to ambiguity in claim-evidence alignment. This issue is further exacerbated by high token-level similarity among related terms, making it difficult for models to differentiate between general and specific concepts. As a result, models often misinterpret evidence relevance, increasing the likelihood of incorrect verification outcomes.

This issue arises when a claim uses a broad term, while the supporting evidence provides a more specific instance, or vice versa. As in the following example, such mismatches can lead to incorrect veracity assignments, as existing models struggle to recognise hierarchical relationships between concepts. 

Claim: \textbf{Cancer} risk is lower in individuals with a history of alcohol consumption. \\
\indent Supports: Alcohol consumption was associated with a decreased risk of \textbf{thyroid cancer}. 

This issue is common within scientific fact checking and has been reported to occur within 44\% of annotated examples in SciFact-Open \cite{wadden-etal-2022-scifact}.
Hence, we argue that capturing the hierarchical relationship could be a research direction to improve verification performance in the scientific domain, by solving the mismatch problem.

\subsection*{\textit{\normalsize{{RD.8. Hierarchical Concept Modelling}}}}
To address this issue, ontology-based reasoning can be integrated into fact-checking pipelines. Structured ontologies such as MeSH  \cite{lipscomb2000medical} and UMLS  \cite{lindberg1993unified} define hierarchical relationships that help systems infer term specificity. Recognising that ‘lung cancer’ is a subtype of ‘cancer’ enables better claim-evidence alignment, mitigating errors caused by lexical similarity. Beyond that, the knowledge graph can enrich ontological reasoning by encoding both hierarchical and associative relationships among scientific concepts  \cite{hoelscher-obermaier-etal-2022-leveraging,kim-etal-2023-factkg}. However, its potential for resolving terminology granularity mismatches in scientific fact-checking remains unexplored. Additionally, representation learning techniques such as contrastive learning can embed these hierarchical relationships into vector space representations, reducing reliance on token-level similarity. Domain-specific models like SciBERT  \cite{beltagy-etal-2019-scibert} and PubMedBERT  \cite{pubmedbert} can further enhance contextual understanding by incorporating structured knowledge into retrieval and verification processes. By leveraging ontological reasoning, knowledge graphs, and structured embeddings, scientific fact-checking systems can better align claims with relevant evidence, reducing verification errors caused by terminology granularity mismatches. In addition to enhancing precision, this also has potential to improve interpretability by making model decisions more transparent.

\section{Conclusion}
This paper explores the evolution of scientific fact-checking methodologies from abstract-level approaches to full-paper frameworks on large-scale corpora. As the volume and diversity of scientific knowledge continue to grow, the challenges of verifying claims across heterogeneous sources become increasingly complex. By addressing the complexities inherent in scientific literature, including its evolving nature, structured format, and the necessity for precise evidence retrieval, we underscore the importance of developing specialised retrieval systems capable of managing large, multimodal, and time-sensitive evidence retrieval. Furthermore, this work proposes several research directions aimed at improving scientific fact-checking efficiency and reliability. They include the integration of time-aware evidence retrieval to ensure the use of the most relevant and up-to-date findings, adaptive document processing to enable context-sensitive retrieval strategies, and multi-modal evidence alignment to integrate text, tables and figures to enhance verification accuracy. Overcoming these challenges has potential to improve the accuracy of fact-checking processes and also facilitate the scalability and applicability of these systems in diverse scientific fact-checking scenarios. By bridging the gap between scientific fact-checking and effective evidence retrieval, these advancements will contribute to more robust, interpretable, and trustworthy scientific fact-checking methodologies for real-world applications.

\bibliographystyle{ACM-Reference-Format}
\bibliography{sample-base}


\begin{thebibliography}{120}


\ifx \showCODEN    \undefined \def \showCODEN     #1{\unskip}     \fi
\ifx \showISBNx    \undefined \def \showISBNx     #1{\unskip}     \fi
\ifx \showISBNxiii \undefined \def \showISBNxiii  #1{\unskip}     \fi
\ifx \showISSN     \undefined \def \showISSN      #1{\unskip}     \fi
\ifx \showLCCN     \undefined \def \showLCCN      #1{\unskip}     \fi
\ifx \shownote     \undefined \def \shownote      #1{#1}          \fi
\ifx \showarticletitle \undefined \def \showarticletitle #1{#1}   \fi
\ifx \showURL      \undefined \def \showURL       {\relax}        \fi
\providecommand\bibfield[2]{#2}
\providecommand\bibinfo[2]{#2}
\providecommand\natexlab[1]{#1}
\providecommand\showeprint[2][]{arXiv:#2}

\bibitem[Abdelnabi et~al\mbox{.}(2022)]%
        {abdelnabi2022open}
\bibfield{author}{\bibinfo{person}{Sahar Abdelnabi}, \bibinfo{person}{Rakibul Hasan}, {and} \bibinfo{person}{Mario Fritz}.} \bibinfo{year}{2022}\natexlab{}.
\newblock \showarticletitle{Open-domain, content-based, multi-modal fact-checking of out-of-context images via online resources}. In \bibinfo{booktitle}{\emph{Proceedings of the IEEE/CVF conference on computer vision and pattern recognition}}. \bibinfo{pages}{14940--14949}.
\newblock


\bibitem[Adlakha et~al\mbox{.}(2024)]%
        {adlakha-etal-2024-evaluating}
\bibfield{author}{\bibinfo{person}{Vaibhav Adlakha}, \bibinfo{person}{Parishad BehnamGhader}, \bibinfo{person}{Xing~Han Lu}, \bibinfo{person}{Nicholas Meade}, {and} \bibinfo{person}{Siva Reddy}.} \bibinfo{year}{2024}\natexlab{}.
\newblock \showarticletitle{Evaluating Correctness and Faithfulness of Instruction-Following Models for Question Answering}.
\newblock \bibinfo{journal}{\emph{Transactions of the Association for Computational Linguistics}}  \bibinfo{volume}{12} (\bibinfo{year}{2024}), \bibinfo{pages}{681--699}.
\newblock
\href{https://doi.org/10.1162/tacl_a_00667}{doi:\nolinkurl{10.1162/tacl_a_00667}}


\bibitem[Akhtar et~al\mbox{.}(2022)]%
        {akhtar-etal-2022-pubhealthtab}
\bibfield{author}{\bibinfo{person}{Mubashara Akhtar}, \bibinfo{person}{Oana Cocarascu}, {and} \bibinfo{person}{Elena Simperl}.} \bibinfo{year}{2022}\natexlab{}.
\newblock \showarticletitle{{P}ub{H}ealth{T}ab: {A} Public Health Table-based Dataset for Evidence-based Fact Checking}. In \bibinfo{booktitle}{\emph{Findings of the Association for Computational Linguistics: NAACL 2022}}, \bibfield{editor}{\bibinfo{person}{Marine Carpuat}, \bibinfo{person}{Marie-Catherine de~Marneffe}, {and} \bibinfo{person}{Ivan~Vladimir Meza~Ruiz}} (Eds.). \bibinfo{publisher}{Association for Computational Linguistics}, \bibinfo{address}{Seattle, United States}, \bibinfo{pages}{1--16}.
\newblock
\href{https://doi.org/10.18653/v1/2022.findings-naacl.1}{doi:\nolinkurl{10.18653/v1/2022.findings-naacl.1}}


\bibitem[Akhtar et~al\mbox{.}(2023a)]%
        {akhtar-etal-2023-reading}
\bibfield{author}{\bibinfo{person}{Mubashara Akhtar}, \bibinfo{person}{Oana Cocarascu}, {and} \bibinfo{person}{Elena Simperl}.} \bibinfo{year}{2023}\natexlab{a}.
\newblock \showarticletitle{Reading and Reasoning over Chart Images for Evidence-based Automated Fact-Checking}. In \bibinfo{booktitle}{\emph{Findings of the Association for Computational Linguistics: EACL 2023}}, \bibfield{editor}{\bibinfo{person}{Andreas Vlachos} {and} \bibinfo{person}{Isabelle Augenstein}} (Eds.). \bibinfo{publisher}{Association for Computational Linguistics}, \bibinfo{address}{Dubrovnik, Croatia}, \bibinfo{pages}{399--414}.
\newblock
\href{https://doi.org/10.18653/v1/2023.findings-eacl.30}{doi:\nolinkurl{10.18653/v1/2023.findings-eacl.30}}


\bibitem[Akhtar et~al\mbox{.}(2023b)]%
        {akhtar-etal-2023-multimodal}
\bibfield{author}{\bibinfo{person}{Mubashara Akhtar}, \bibinfo{person}{Michael Schlichtkrull}, \bibinfo{person}{Zhijiang Guo}, \bibinfo{person}{Oana Cocarascu}, \bibinfo{person}{Elena Simperl}, {and} \bibinfo{person}{Andreas Vlachos}.} \bibinfo{year}{2023}\natexlab{b}.
\newblock \showarticletitle{Multimodal Automated Fact-Checking: A Survey}. In \bibinfo{booktitle}{\emph{Findings of the Association for Computational Linguistics: EMNLP 2023}}, \bibfield{editor}{\bibinfo{person}{Houda Bouamor}, \bibinfo{person}{Juan Pino}, {and} \bibinfo{person}{Kalika Bali}} (Eds.). \bibinfo{publisher}{Association for Computational Linguistics}, \bibinfo{address}{Singapore}, \bibinfo{pages}{5430--5448}.
\newblock
\href{https://doi.org/10.18653/v1/2023.findings-emnlp.361}{doi:\nolinkurl{10.18653/v1/2023.findings-emnlp.361}}


\bibitem[Akhtar et~al\mbox{.}(2024a)]%
        {akhtar2024ev2r}
\bibfield{author}{\bibinfo{person}{Mubashara Akhtar}, \bibinfo{person}{Michael Schlichtkrull}, {and} \bibinfo{person}{Andreas Vlachos}.} \bibinfo{year}{2024}\natexlab{a}.
\newblock \showarticletitle{Ev2R: Evaluating Evidence Retrieval in Automated Fact-Checking}.
\newblock \bibinfo{journal}{\emph{arXiv preprint arXiv:2411.05375}} (\bibinfo{year}{2024}).
\newblock


\bibitem[Akhtar et~al\mbox{.}(2024b)]%
        {akhtar-etal-2024-chartcheck}
\bibfield{author}{\bibinfo{person}{Mubashara Akhtar}, \bibinfo{person}{Nikesh Subedi}, \bibinfo{person}{Vivek Gupta}, \bibinfo{person}{Sahar Tahmasebi}, \bibinfo{person}{Oana Cocarascu}, {and} \bibinfo{person}{Elena Simperl}.} \bibinfo{year}{2024}\natexlab{b}.
\newblock \showarticletitle{{C}hart{C}heck: Explainable Fact-Checking over Real-World Chart Images}. In \bibinfo{booktitle}{\emph{Findings of the Association for Computational Linguistics: ACL 2024}}, \bibfield{editor}{\bibinfo{person}{Lun-Wei Ku}, \bibinfo{person}{Andre Martins}, {and} \bibinfo{person}{Vivek Srikumar}} (Eds.). \bibinfo{publisher}{Association for Computational Linguistics}, \bibinfo{address}{Bangkok, Thailand}, \bibinfo{pages}{13921--13937}.
\newblock
\href{https://doi.org/10.18653/v1/2024.findings-acl.828}{doi:\nolinkurl{10.18653/v1/2024.findings-acl.828}}


\bibitem[Allein et~al\mbox{.}(2023)]%
        {allein-etal-2023-implicit}
\bibfield{author}{\bibinfo{person}{Liesbeth Allein}, \bibinfo{person}{Marlon Saelens}, \bibinfo{person}{Ruben Cartuyvels}, {and} \bibinfo{person}{Marie-Francine Moens}.} \bibinfo{year}{2023}\natexlab{}.
\newblock \showarticletitle{Implicit Temporal Reasoning for Evidence-Based Fact-Checking}. In \bibinfo{booktitle}{\emph{Findings of the Association for Computational Linguistics: EACL 2023}}, \bibfield{editor}{\bibinfo{person}{Andreas Vlachos} {and} \bibinfo{person}{Isabelle Augenstein}} (Eds.). \bibinfo{publisher}{Association for Computational Linguistics}, \bibinfo{address}{Dubrovnik, Croatia}, \bibinfo{pages}{176--189}.
\newblock
\href{https://doi.org/10.18653/v1/2023.findings-eacl.13}{doi:\nolinkurl{10.18653/v1/2023.findings-eacl.13}}


\bibitem[Augenstein et~al\mbox{.}(2019)]%
        {augenstein-etal-2019-multifc}
\bibfield{author}{\bibinfo{person}{Isabelle Augenstein}, \bibinfo{person}{Christina Lioma}, \bibinfo{person}{Dongsheng Wang}, \bibinfo{person}{Lucas Chaves~Lima}, \bibinfo{person}{Casper Hansen}, \bibinfo{person}{Christian Hansen}, {and} \bibinfo{person}{Jakob~Grue Simonsen}.} \bibinfo{year}{2019}\natexlab{}.
\newblock \showarticletitle{{M}ulti{FC}: A Real-World Multi-Domain Dataset for Evidence-Based Fact Checking of Claims}. In \bibinfo{booktitle}{\emph{Proceedings of the 2019 Conference on Empirical Methods in Natural Language Processing and the 9th International Joint Conference on Natural Language Processing (EMNLP-IJCNLP)}}, \bibfield{editor}{\bibinfo{person}{Kentaro Inui}, \bibinfo{person}{Jing Jiang}, \bibinfo{person}{Vincent Ng}, {and} \bibinfo{person}{Xiaojun Wan}} (Eds.). \bibinfo{publisher}{Association for Computational Linguistics}, \bibinfo{address}{Hong Kong, China}, \bibinfo{pages}{4685--4697}.
\newblock
\href{https://doi.org/10.18653/v1/D19-1475}{doi:\nolinkurl{10.18653/v1/D19-1475}}


\bibitem[Bahri et~al\mbox{.}(2020)]%
        {bahri2020choppy}
\bibfield{author}{\bibinfo{person}{Dara Bahri}, \bibinfo{person}{Yi Tay}, \bibinfo{person}{Che Zheng}, \bibinfo{person}{Donald Metzler}, {and} \bibinfo{person}{Andrew Tomkins}.} \bibinfo{year}{2020}\natexlab{}.
\newblock \showarticletitle{Choppy: Cut transformer for ranked list truncation}. In \bibinfo{booktitle}{\emph{Proceedings of the 43rd International ACM SIGIR Conference on Research and Development in Information Retrieval}}. \bibinfo{pages}{1513--1516}.
\newblock


\bibitem[Beltagy et~al\mbox{.}(2019)]%
        {beltagy-etal-2019-scibert}
\bibfield{author}{\bibinfo{person}{Iz Beltagy}, \bibinfo{person}{Kyle Lo}, {and} \bibinfo{person}{Arman Cohan}.} \bibinfo{year}{2019}\natexlab{}.
\newblock \showarticletitle{{S}ci{BERT}: A Pretrained Language Model for Scientific Text}. In \bibinfo{booktitle}{\emph{Proceedings of the 2019 Conference on Empirical Methods in Natural Language Processing and the 9th International Joint Conference on Natural Language Processing (EMNLP-IJCNLP)}}, \bibfield{editor}{\bibinfo{person}{Kentaro Inui}, \bibinfo{person}{Jing Jiang}, \bibinfo{person}{Vincent Ng}, {and} \bibinfo{person}{Xiaojun Wan}} (Eds.). \bibinfo{publisher}{Association for Computational Linguistics}, \bibinfo{address}{Hong Kong, China}, \bibinfo{pages}{3615--3620}.
\newblock
\href{https://doi.org/10.18653/v1/D19-1371}{doi:\nolinkurl{10.18653/v1/D19-1371}}


\bibitem[Beltagy et~al\mbox{.}(2020)]%
        {Beltagy2020Longformer}
\bibfield{author}{\bibinfo{person}{Iz Beltagy}, \bibinfo{person}{Matthew~E. Peters}, {and} \bibinfo{person}{Arman Cohan}.} \bibinfo{year}{2020}\natexlab{}.
\newblock \showarticletitle{Longformer: The Long-Document Transformer}.
\newblock \bibinfo{journal}{\emph{arXiv:2004.05150}} (\bibinfo{year}{2020}).
\newblock


\bibitem[Bhagavatula et~al\mbox{.}(2013)]%
        {bhagavatula2013methods}
\bibfield{author}{\bibinfo{person}{Chandra~Sekhar Bhagavatula}, \bibinfo{person}{Thanapon Noraset}, {and} \bibinfo{person}{Doug Downey}.} \bibinfo{year}{2013}\natexlab{}.
\newblock \showarticletitle{Methods for exploring and mining tables on wikipedia}. In \bibinfo{booktitle}{\emph{Proceedings of the ACM SIGKDD workshop on interactive data exploration and analytics}}. \bibinfo{pages}{18--26}.
\newblock


\bibitem[Bin-Hezam and Stevenson(2023)]%
        {hezam-stevenson-2023-combining}
\bibfield{author}{\bibinfo{person}{Reem Bin-Hezam} {and} \bibinfo{person}{Mark Stevenson}.} \bibinfo{year}{2023}\natexlab{}.
\newblock \showarticletitle{Combining Counting Processes and Classification Improves a Stopping Rule for Technology Assisted Review}. In \bibinfo{booktitle}{\emph{Findings of the Association for Computational Linguistics: EMNLP 2023}}, \bibfield{editor}{\bibinfo{person}{Houda Bouamor}, \bibinfo{person}{Juan Pino}, {and} \bibinfo{person}{Kalika Bali}} (Eds.). \bibinfo{publisher}{Association for Computational Linguistics}, \bibinfo{address}{Singapore}, \bibinfo{pages}{2603--2609}.
\newblock
\href{https://doi.org/10.18653/v1/2023.findings-emnlp.171}{doi:\nolinkurl{10.18653/v1/2023.findings-emnlp.171}}


\bibitem[Bin-Hezam and Stevenson(2024)]%
        {bin2024rlstop}
\bibfield{author}{\bibinfo{person}{Reem Bin-Hezam} {and} \bibinfo{person}{Mark Stevenson}.} \bibinfo{year}{2024}\natexlab{}.
\newblock \showarticletitle{RLStop: A Reinforcement Learning Stopping Method for TAR}. In \bibinfo{booktitle}{\emph{Proceedings of the 47th International ACM SIGIR Conference on Research and Development in Information Retrieval}}. \bibinfo{pages}{2604--2608}.
\newblock


\bibitem[Bulian et~al\mbox{.}(2020)]%
        {bulian2020climate}
\bibfield{author}{\bibinfo{person}{Jannis Bulian}, \bibinfo{person}{Jordan Boyd-Graber}, \bibinfo{person}{Markus Leippold}, \bibinfo{person}{Massimiliano Ciaramita}, {and} \bibinfo{person}{Thomas Diggelmann}.} \bibinfo{year}{2020}\natexlab{}.
\newblock \showarticletitle{Climate-fever: A dataset for verification of real-world climate claims}. In \bibinfo{booktitle}{\emph{NeurIPS 2020 Workshop on Tackling Climate Change with Machine Learning}}.
\newblock


\bibitem[Buneman et~al\mbox{.}(2021)]%
        {buneman2021data}
\bibfield{author}{\bibinfo{person}{Peter Buneman}, \bibinfo{person}{Dennis Dosso}, \bibinfo{person}{Matteo Lissandrini}, {and} \bibinfo{person}{Gianmaria Silvello}.} \bibinfo{year}{2021}\natexlab{}.
\newblock \showarticletitle{Data citation and the citation graph}.
\newblock \bibinfo{journal}{\emph{Quantitative Science Studies}} \bibinfo{volume}{2}, \bibinfo{number}{4} (\bibinfo{year}{2021}), \bibinfo{pages}{1399--1422}.
\newblock


\bibitem[Cao et~al\mbox{.}(2025)]%
        {cao2025averimatecdatasetautomaticverification}
\bibfield{author}{\bibinfo{person}{Rui Cao}, \bibinfo{person}{Zifeng Ding}, \bibinfo{person}{Zhijiang Guo}, \bibinfo{person}{Michael Schlichtkrull}, {and} \bibinfo{person}{Andreas Vlachos}.} \bibinfo{year}{2025}\natexlab{}.
\newblock \bibinfo{title}{AVerImaTeC: A Dataset for Automatic Verification of Image-Text Claims with Evidence from the Web}.
\newblock
\showeprint[arxiv]{2505.17978}~[cs.CL]
\urldef\tempurl%
\url{https://arxiv.org/abs/2505.17978}
\showURL{%
\tempurl}


\bibitem[Chen et~al\mbox{.}(2023)]%
        {chen-etal-2023-layout}
\bibfield{author}{\bibinfo{person}{Catherine Chen}, \bibinfo{person}{Zejiang Shen}, \bibinfo{person}{Dan Klein}, \bibinfo{person}{Gabriel Stanovsky}, \bibinfo{person}{Doug Downey}, {and} \bibinfo{person}{Kyle Lo}.} \bibinfo{year}{2023}\natexlab{}.
\newblock \showarticletitle{Are Layout-Infused Language Models Robust to Layout Distribution Shifts? A Case Study with Scientific Documents}. In \bibinfo{booktitle}{\emph{Findings of the Association for Computational Linguistics: ACL 2023}}, \bibfield{editor}{\bibinfo{person}{Anna Rogers}, \bibinfo{person}{Jordan Boyd-Graber}, {and} \bibinfo{person}{Naoaki Okazaki}} (Eds.). \bibinfo{publisher}{Association for Computational Linguistics}, \bibinfo{address}{Toronto, Canada}, \bibinfo{pages}{13345--13360}.
\newblock
\href{https://doi.org/10.18653/v1/2023.findings-acl.844}{doi:\nolinkurl{10.18653/v1/2023.findings-acl.844}}


\bibitem[Chen et~al\mbox{.}(2024)]%
        {chen2024benchmarking}
\bibfield{author}{\bibinfo{person}{Jiawei Chen}, \bibinfo{person}{Hongyu Lin}, \bibinfo{person}{Xianpei Han}, {and} \bibinfo{person}{Le Sun}.} \bibinfo{year}{2024}\natexlab{}.
\newblock \showarticletitle{Benchmarking large language models in retrieval-augmented generation}. In \bibinfo{booktitle}{\emph{Proceedings of the AAAI Conference on Artificial Intelligence}}, Vol.~\bibinfo{volume}{38}. \bibinfo{pages}{17754--17762}.
\newblock


\bibitem[Chen et~al\mbox{.}(2019)]%
        {chen2019biosentvec}
\bibfield{author}{\bibinfo{person}{Qingyu Chen}, \bibinfo{person}{Yifan Peng}, {and} \bibinfo{person}{Zhiyong Lu}.} \bibinfo{year}{2019}\natexlab{}.
\newblock \showarticletitle{BioSentVec: creating sentence embeddings for biomedical texts}. In \bibinfo{booktitle}{\emph{2019 IEEE International Conference on Healthcare Informatics (ICHI)}}. IEEE, \bibinfo{pages}{1--5}.
\newblock


\bibitem[Chen et~al\mbox{.}(2020)]%
        {2019TabFactA}
\bibfield{author}{\bibinfo{person}{Wenhu Chen}, \bibinfo{person}{Hongmin Wang}, \bibinfo{person}{Jianshu Chen}, \bibinfo{person}{Yunkai Zhang}, \bibinfo{person}{Hong Wang}, \bibinfo{person}{Shiyang Li}, \bibinfo{person}{Xiyou Zhou}, {and} \bibinfo{person}{William~Yang Wang}.} \bibinfo{year}{2020}\natexlab{}.
\newblock \showarticletitle{TabFact : A Large-scale Dataset for Table-based Fact Verification}. In \bibinfo{booktitle}{\emph{International Conference on Learning Representations (ICLR)}}. \bibinfo{address}{Addis Ababa, Ethiopia}.
\newblock


\bibitem[Chiesurin et~al\mbox{.}(2023)]%
        {chiesurin-etal-2023-dangers}
\bibfield{author}{\bibinfo{person}{Sabrina Chiesurin}, \bibinfo{person}{Dimitris Dimakopoulos}, \bibinfo{person}{Marco~Antonio Sobrevilla~Cabezudo}, \bibinfo{person}{Arash Eshghi}, \bibinfo{person}{Ioannis Papaioannou}, \bibinfo{person}{Verena Rieser}, {and} \bibinfo{person}{Ioannis Konstas}.} \bibinfo{year}{2023}\natexlab{}.
\newblock \showarticletitle{The Dangers of trusting Stochastic Parrots: Faithfulness and Trust in Open-domain Conversational Question Answering}. In \bibinfo{booktitle}{\emph{Findings of the Association for Computational Linguistics: ACL 2023}}, \bibfield{editor}{\bibinfo{person}{Anna Rogers}, \bibinfo{person}{Jordan Boyd-Graber}, {and} \bibinfo{person}{Naoaki Okazaki}} (Eds.). \bibinfo{publisher}{Association for Computational Linguistics}, \bibinfo{address}{Toronto, Canada}, \bibinfo{pages}{947--959}.
\newblock
\href{https://doi.org/10.18653/v1/2023.findings-acl.60}{doi:\nolinkurl{10.18653/v1/2023.findings-acl.60}}


\bibitem[Cho et~al\mbox{.}(2024)]%
        {cho2024m3docrag}
\bibfield{author}{\bibinfo{person}{Jaemin Cho}, \bibinfo{person}{Debanjan Mahata}, \bibinfo{person}{Ozan Irsoy}, \bibinfo{person}{Yujie He}, {and} \bibinfo{person}{Mohit Bansal}.} \bibinfo{year}{2024}\natexlab{}.
\newblock \showarticletitle{M3docrag: Multi-modal retrieval is what you need for multi-page multi-document understanding}.
\newblock \bibinfo{journal}{\emph{arXiv preprint arXiv:2411.04952}} (\bibinfo{year}{2024}).
\newblock


\bibitem[Cormack and Grossman(2014)]%
        {cormack2014evaluation}
\bibfield{author}{\bibinfo{person}{Gordon~V Cormack} {and} \bibinfo{person}{Maura~R Grossman}.} \bibinfo{year}{2014}\natexlab{}.
\newblock \showarticletitle{Evaluation of machine-learning protocols for technology-assisted review in electronic discovery}. In \bibinfo{booktitle}{\emph{Proceedings of the 37th international ACM SIGIR conference on Research \& development in information retrieval}}. \bibinfo{pages}{153--162}.
\newblock


\bibitem[Craswell et~al\mbox{.}(2021)]%
        {craswell2021overviewtrec2020deep}
\bibfield{author}{\bibinfo{person}{Nick Craswell}, \bibinfo{person}{Bhaskar Mitra}, \bibinfo{person}{Emine Yilmaz}, {and} \bibinfo{person}{Daniel Campos}.} \bibinfo{year}{2021}\natexlab{}.
\newblock \bibinfo{title}{Overview of the TREC 2020 deep learning track}.
\newblock
\showeprint[arxiv]{2102.07662}~[cs.IR]
\urldef\tempurl%
\url{https://arxiv.org/abs/2102.07662}
\showURL{%
\tempurl}


\bibitem[Craswell et~al\mbox{.}(2020)]%
        {craswell2020overview}
\bibfield{author}{\bibinfo{person}{Nick Craswell}, \bibinfo{person}{Bhaskar Mitra}, \bibinfo{person}{Emine Yilmaz}, \bibinfo{person}{Daniel Campos}, {and} \bibinfo{person}{Ellen~M Voorhees}.} \bibinfo{year}{2020}\natexlab{}.
\newblock \showarticletitle{Overview of the TREC 2019 deep learning track}.
\newblock \bibinfo{journal}{\emph{arXiv preprint arXiv:2003.07820}} (\bibinfo{year}{2020}).
\newblock


\bibitem[Dmonte et~al\mbox{.}(2024)]%
        {dmonte2024claim}
\bibfield{author}{\bibinfo{person}{Alphaeus Dmonte}, \bibinfo{person}{Roland Oruche}, \bibinfo{person}{Marcos Zampieri}, \bibinfo{person}{Prasad Calyam}, {and} \bibinfo{person}{Isabelle Augenstein}.} \bibinfo{year}{2024}\natexlab{}.
\newblock \showarticletitle{Claim Verification in the Age of Large Language Models: A Survey}.
\newblock \bibinfo{journal}{\emph{arXiv preprint arXiv:2408.14317}} (\bibinfo{year}{2024}).
\newblock


\bibitem[Dong and Wang(2024)]%
        {dong2024large}
\bibfield{author}{\bibinfo{person}{Haoyu Dong} {and} \bibinfo{person}{Zhiruo Wang}.} \bibinfo{year}{2024}\natexlab{}.
\newblock \showarticletitle{Large language models for tabular data: Progresses and future directions}. In \bibinfo{booktitle}{\emph{Proceedings of the 47th International ACM SIGIR Conference on Research and Development in Information Retrieval}}. \bibinfo{pages}{2997--3000}.
\newblock


\bibitem[F{\"a}rber and Jatowt(2020)]%
        {farber2020citation}
\bibfield{author}{\bibinfo{person}{Michael F{\"a}rber} {and} \bibinfo{person}{Adam Jatowt}.} \bibinfo{year}{2020}\natexlab{}.
\newblock \showarticletitle{Citation recommendation: approaches and datasets}.
\newblock \bibinfo{journal}{\emph{International Journal on Digital Libraries}} \bibinfo{volume}{21}, \bibinfo{number}{4} (\bibinfo{year}{2020}), \bibinfo{pages}{375--405}.
\newblock


\bibitem[Fletcher and Stevenson(2025)]%
        {fletcher2025predicting}
\bibfield{author}{\bibinfo{person}{Aaron~HA Fletcher} {and} \bibinfo{person}{Mark Stevenson}.} \bibinfo{year}{2025}\natexlab{}.
\newblock \showarticletitle{Predicting retracted research: a dataset and machine learning approaches}.
\newblock \bibinfo{journal}{\emph{Research Integrity and Peer Review}} \bibinfo{volume}{10}, \bibinfo{number}{1} (\bibinfo{year}{2025}), \bibinfo{pages}{1--10}.
\newblock


\bibitem[Fok et~al\mbox{.}(2023)]%
        {fok2023scim}
\bibfield{author}{\bibinfo{person}{Raymond Fok}, \bibinfo{person}{Hita Kambhamettu}, \bibinfo{person}{Luca Soldaini}, \bibinfo{person}{Jonathan Bragg}, \bibinfo{person}{Kyle Lo}, \bibinfo{person}{Marti Hearst}, \bibinfo{person}{Andrew Head}, {and} \bibinfo{person}{Daniel~S Weld}.} \bibinfo{year}{2023}\natexlab{}.
\newblock \showarticletitle{Scim: Intelligent skimming support for scientific papers}. In \bibinfo{booktitle}{\emph{Proceedings of the 28th International Conference on Intelligent User Interfaces}}. \bibinfo{pages}{476--490}.
\newblock


\bibitem[Grossman et~al\mbox{.}(2016)]%
        {grossman2016trec}
\bibfield{author}{\bibinfo{person}{Maura~R Grossman}, \bibinfo{person}{Gordon~V Cormack}, {and} \bibinfo{person}{Adam Roegiest}.} \bibinfo{year}{2016}\natexlab{}.
\newblock \showarticletitle{TREC 2016 Total Recall Track Overview.}. In \bibinfo{booktitle}{\emph{TREC}}.
\newblock


\bibitem[Gu et~al\mbox{.}(2022b)]%
        {gu2022local}
\bibfield{author}{\bibinfo{person}{Nianlong Gu}, \bibinfo{person}{Yingqiang Gao}, {and} \bibinfo{person}{Richard~HR Hahnloser}.} \bibinfo{year}{2022}\natexlab{b}.
\newblock \showarticletitle{Local citation recommendation with hierarchical-attention text encoder and scibert-based reranking}. In \bibinfo{booktitle}{\emph{European conference on information retrieval}}. Springer, \bibinfo{pages}{274--288}.
\newblock


\bibitem[Gu et~al\mbox{.}(2020)]%
        {pubmedbert}
\bibfield{author}{\bibinfo{person}{Yu Gu}, \bibinfo{person}{Robert Tinn}, \bibinfo{person}{Hao Cheng}, \bibinfo{person}{Michael Lucas}, \bibinfo{person}{Naoto Usuyama}, \bibinfo{person}{Xiaodong Liu}, \bibinfo{person}{Tristan Naumann}, \bibinfo{person}{Jianfeng Gao}, {and} \bibinfo{person}{Hoifung Poon}.} \bibinfo{year}{2020}\natexlab{}.
\newblock \bibinfo{title}{Domain-Specific Language Model Pretraining for Biomedical Natural Language Processing}.
\newblock
\showeprint{arXiv:2007.15779}


\bibitem[Gu et~al\mbox{.}(2022a)]%
        {gu-etal-2022-pasta}
\bibfield{author}{\bibinfo{person}{Zihui Gu}, \bibinfo{person}{Ju Fan}, \bibinfo{person}{Nan Tang}, \bibinfo{person}{Preslav Nakov}, \bibinfo{person}{Xiaoman Zhao}, {and} \bibinfo{person}{Xiaoyong Du}.} \bibinfo{year}{2022}\natexlab{a}.
\newblock \showarticletitle{{PASTA}: Table-Operations Aware Fact Verification via Sentence-Table Cloze Pre-training}. In \bibinfo{booktitle}{\emph{Proceedings of the 2022 Conference on Empirical Methods in Natural Language Processing}}, \bibfield{editor}{\bibinfo{person}{Yoav Goldberg}, \bibinfo{person}{Zornitsa Kozareva}, {and} \bibinfo{person}{Yue Zhang}} (Eds.). \bibinfo{publisher}{Association for Computational Linguistics}, \bibinfo{address}{Abu Dhabi, United Arab Emirates}, \bibinfo{pages}{4971--4983}.
\newblock
\href{https://doi.org/10.18653/v1/2022.emnlp-main.331}{doi:\nolinkurl{10.18653/v1/2022.emnlp-main.331}}


\bibitem[Guo et~al\mbox{.}(2022)]%
        {guo2022survey}
\bibfield{author}{\bibinfo{person}{Zhijiang Guo}, \bibinfo{person}{Michael Schlichtkrull}, {and} \bibinfo{person}{Andreas Vlachos}.} \bibinfo{year}{2022}\natexlab{}.
\newblock \showarticletitle{A survey on automated fact-checking}.
\newblock \bibinfo{journal}{\emph{Transactions of the Association for Computational Linguistics}}  \bibinfo{volume}{10} (\bibinfo{year}{2022}), \bibinfo{pages}{178--206}.
\newblock


\bibitem[Herzig et~al\mbox{.}(2020)]%
        {herzig-etal-2020-tapas}
\bibfield{author}{\bibinfo{person}{Jonathan Herzig}, \bibinfo{person}{Pawel~Krzysztof Nowak}, \bibinfo{person}{Thomas M{\"u}ller}, \bibinfo{person}{Francesco Piccinno}, {and} \bibinfo{person}{Julian Eisenschlos}.} \bibinfo{year}{2020}\natexlab{}.
\newblock \showarticletitle{{T}a{P}as: Weakly Supervised Table Parsing via Pre-training}. In \bibinfo{booktitle}{\emph{Proceedings of the 58th Annual Meeting of the Association for Computational Linguistics}}, \bibfield{editor}{\bibinfo{person}{Dan Jurafsky}, \bibinfo{person}{Joyce Chai}, \bibinfo{person}{Natalie Schluter}, {and} \bibinfo{person}{Joel Tetreault}} (Eds.). \bibinfo{publisher}{Association for Computational Linguistics}, \bibinfo{address}{Online}, \bibinfo{pages}{4320--4333}.
\newblock
\href{https://doi.org/10.18653/v1/2020.acl-main.398}{doi:\nolinkurl{10.18653/v1/2020.acl-main.398}}


\bibitem[Hoelscher-Obermaier et~al\mbox{.}(2022)]%
        {hoelscher-obermaier-etal-2022-leveraging}
\bibfield{author}{\bibinfo{person}{Jason Hoelscher-Obermaier}, \bibinfo{person}{Edward Stevinson}, \bibinfo{person}{Valentin Stauber}, \bibinfo{person}{Ivaylo Zhelev}, \bibinfo{person}{Viktor Botev}, \bibinfo{person}{Ronin Wu}, {and} \bibinfo{person}{Jeremy Minton}.} \bibinfo{year}{2022}\natexlab{}.
\newblock \showarticletitle{Leveraging knowledge graphs to update scientific word embeddings using latent semantic imputation}. In \bibinfo{booktitle}{\emph{Proceedings of the first Workshop on Information Extraction from Scientific Publications}}, \bibfield{editor}{\bibinfo{person}{Tirthankar Ghosal}, \bibinfo{person}{Sergi Blanco-Cuaresma}, \bibinfo{person}{Alberto Accomazzi}, \bibinfo{person}{Robert~M. Patton}, \bibinfo{person}{Felix Grezes}, {and} \bibinfo{person}{Thomas Allen}} (Eds.). \bibinfo{publisher}{Association for Computational Linguistics}, \bibinfo{address}{Online}, \bibinfo{pages}{43--53}.
\newblock
\href{https://doi.org/10.18653/v1/2022.wiesp-1.6}{doi:\nolinkurl{10.18653/v1/2022.wiesp-1.6}}


\bibitem[Hu et~al\mbox{.}(2023)]%
        {hu2023read}
\bibfield{author}{\bibinfo{person}{Xuming Hu}, \bibinfo{person}{Zhaochen Hong}, \bibinfo{person}{Zhijiang Guo}, \bibinfo{person}{Lijie Wen}, {and} \bibinfo{person}{Philip Yu}.} \bibinfo{year}{2023}\natexlab{}.
\newblock \showarticletitle{Read it twice: Towards faithfully interpretable fact verification by revisiting evidence}. In \bibinfo{booktitle}{\emph{Proceedings of the 46th International ACM SIGIR Conference on Research and Development in Information Retrieval}}. \bibinfo{pages}{2319--2323}.
\newblock


\bibitem[Huang et~al\mbox{.}(2022)]%
        {huang-etal-2022-lightweight}
\bibfield{author}{\bibinfo{person}{Po-Wei Huang}, \bibinfo{person}{Abhinav Ramesh~Kashyap}, \bibinfo{person}{Yanxia Qin}, \bibinfo{person}{Yajing Yang}, {and} \bibinfo{person}{Min-Yen Kan}.} \bibinfo{year}{2022}\natexlab{}.
\newblock \showarticletitle{Lightweight Contextual Logical Structure Recovery}. In \bibinfo{booktitle}{\emph{Proceedings of the Third Workshop on Scholarly Document Processing}}, \bibfield{editor}{\bibinfo{person}{Arman Cohan}, \bibinfo{person}{Guy Feigenblat}, \bibinfo{person}{Dayne Freitag}, \bibinfo{person}{Tirthankar Ghosal}, \bibinfo{person}{Drahomira Herrmannova}, \bibinfo{person}{Petr Knoth}, \bibinfo{person}{Kyle Lo}, \bibinfo{person}{Philipp Mayr}, \bibinfo{person}{Michal Shmueli-Scheuer}, \bibinfo{person}{Anita de~Waard}, {and} \bibinfo{person}{Lucy~Lu Wang}} (Eds.). \bibinfo{publisher}{Association for Computational Linguistics}, \bibinfo{address}{Gyeongju, Republic of Korea}, \bibinfo{pages}{37--48}.
\newblock
\urldef\tempurl%
\url{https://aclanthology.org/2022.sdp-1.5/}
\showURL{%
\tempurl}


\bibitem[Ikoma and Matsubara(2023)]%
        {ikoma-matsubara-2023-use}
\bibfield{author}{\bibinfo{person}{Tomoki Ikoma} {and} \bibinfo{person}{Shigeki Matsubara}.} \bibinfo{year}{2023}\natexlab{}.
\newblock \showarticletitle{On the Use of Language Models for Function Identification of Citations in Scholarly Papers}. In \bibinfo{booktitle}{\emph{Proceedings of the Second Workshop on Information Extraction from Scientific Publications}}, \bibfield{editor}{\bibinfo{person}{Tirthankar Ghosal}, \bibinfo{person}{Felix Grezes}, \bibinfo{person}{Thomas Allen}, \bibinfo{person}{Kelly Lockhart}, \bibinfo{person}{Alberto Accomazzi}, {and} \bibinfo{person}{Sergi Blanco-Cuaresma}} (Eds.). \bibinfo{publisher}{Association for Computational Linguistics}, \bibinfo{address}{Bali, Indonesia}, \bibinfo{pages}{130--135}.
\newblock
\href{https://doi.org/10.18653/v1/2023.wiesp-1.15}{doi:\nolinkurl{10.18653/v1/2023.wiesp-1.15}}


\bibitem[Kanoulas et~al\mbox{.}(2018a)]%
        {Kanoulas2018CLEF2T}
\bibfield{author}{\bibinfo{person}{E. Kanoulas}, \bibinfo{person}{Dan Li}, \bibinfo{person}{Leif Azzopardi}, {and} \bibinfo{person}{Ren{\'e} Spijker}.} \bibinfo{year}{2018}\natexlab{a}.
\newblock \showarticletitle{CLEF 2017 Technologically Assisted Reviews in Empirical Medicine Overview}. In \bibinfo{booktitle}{\emph{Conference and Labs of the Evaluation Forum}}.
\newblock
\urldef\tempurl%
\url{https://api.semanticscholar.org/CorpusID:246440739}
\showURL{%
\tempurl}


\bibitem[Kanoulas et~al\mbox{.}(2018b)]%
        {kanoulas2018clef}
\bibfield{author}{\bibinfo{person}{Evangelos Kanoulas}, \bibinfo{person}{Dan Li}, \bibinfo{person}{Leif Azzopardi}, {and} \bibinfo{person}{Rene Spijker}.} \bibinfo{year}{2018}\natexlab{b}.
\newblock \showarticletitle{CLEF 2018 technologically assisted reviews in empirical medicine overview}. In \bibinfo{booktitle}{\emph{CEUR workshop proceedings}}, Vol.~\bibinfo{volume}{2125}.
\newblock


\bibitem[Kanoulas et~al\mbox{.}(2019)]%
        {kanoulas2019clef}
\bibfield{author}{\bibinfo{person}{Evangelos Kanoulas}, \bibinfo{person}{Dan Li}, \bibinfo{person}{Leif Azzopardi}, {and} \bibinfo{person}{Rene Spijker}.} \bibinfo{year}{2019}\natexlab{}.
\newblock \showarticletitle{CLEF 2019 technology assisted reviews in empirical medicine overview}. In \bibinfo{booktitle}{\emph{CEUR workshop proceedings}}, Vol.~\bibinfo{volume}{2380}. \bibinfo{pages}{250}.
\newblock


\bibitem[Karpinska et~al\mbox{.}(2024)]%
        {karpinska-etal-2024-one}
\bibfield{author}{\bibinfo{person}{Marzena Karpinska}, \bibinfo{person}{Katherine Thai}, \bibinfo{person}{Kyle Lo}, \bibinfo{person}{Tanya Goyal}, {and} \bibinfo{person}{Mohit Iyyer}.} \bibinfo{year}{2024}\natexlab{}.
\newblock \showarticletitle{One Thousand and One Pairs: A {``}novel{''} challenge for long-context language models}. In \bibinfo{booktitle}{\emph{Proceedings of the 2024 Conference on Empirical Methods in Natural Language Processing}}, \bibfield{editor}{\bibinfo{person}{Yaser Al-Onaizan}, \bibinfo{person}{Mohit Bansal}, {and} \bibinfo{person}{Yun-Nung Chen}} (Eds.). \bibinfo{publisher}{Association for Computational Linguistics}, \bibinfo{address}{Miami, Florida, USA}, \bibinfo{pages}{17048--17085}.
\newblock
\href{https://doi.org/10.18653/v1/2024.emnlp-main.948}{doi:\nolinkurl{10.18653/v1/2024.emnlp-main.948}}


\bibitem[Khaliq et~al\mbox{.}(2024)]%
        {khaliq-etal-2024-ragar}
\bibfield{author}{\bibinfo{person}{Mohammed~Abdul Khaliq}, \bibinfo{person}{Paul Yu-Chun Chang}, \bibinfo{person}{Mingyang Ma}, \bibinfo{person}{Bernhard Pflugfelder}, {and} \bibinfo{person}{Filip Mileti{\'c}}.} \bibinfo{year}{2024}\natexlab{}.
\newblock \showarticletitle{{RAGAR}, Your Falsehood Radar: {RAG}-Augmented Reasoning for Political Fact-Checking using Multimodal Large Language Models}. In \bibinfo{booktitle}{\emph{Proceedings of the Seventh Fact Extraction and VERification Workshop (FEVER)}}, \bibfield{editor}{\bibinfo{person}{Michael Schlichtkrull}, \bibinfo{person}{Yulong Chen}, \bibinfo{person}{Chenxi Whitehouse}, \bibinfo{person}{Zhenyun Deng}, \bibinfo{person}{Mubashara Akhtar}, \bibinfo{person}{Rami Aly}, \bibinfo{person}{Zhijiang Guo}, \bibinfo{person}{Christos Christodoulopoulos}, \bibinfo{person}{Oana Cocarascu}, \bibinfo{person}{Arpit Mittal}, \bibinfo{person}{James Thorne}, {and} \bibinfo{person}{Andreas Vlachos}} (Eds.). \bibinfo{publisher}{Association for Computational Linguistics}, \bibinfo{address}{Miami, Florida, USA}, \bibinfo{pages}{280--296}.
\newblock
\href{https://doi.org/10.18653/v1/2024.fever-1.29}{doi:\nolinkurl{10.18653/v1/2024.fever-1.29}}


\bibitem[Kim et~al\mbox{.}(2023)]%
        {kim-etal-2023-factkg}
\bibfield{author}{\bibinfo{person}{Jiho Kim}, \bibinfo{person}{Sungjin Park}, \bibinfo{person}{Yeonsu Kwon}, \bibinfo{person}{Yohan Jo}, \bibinfo{person}{James Thorne}, {and} \bibinfo{person}{Edward Choi}.} \bibinfo{year}{2023}\natexlab{}.
\newblock \showarticletitle{{F}act{KG}: Fact Verification via Reasoning on Knowledge Graphs}. In \bibinfo{booktitle}{\emph{Proceedings of the 61st Annual Meeting of the Association for Computational Linguistics (Volume 1: Long Papers)}}, \bibfield{editor}{\bibinfo{person}{Anna Rogers}, \bibinfo{person}{Jordan Boyd-Graber}, {and} \bibinfo{person}{Naoaki Okazaki}} (Eds.). \bibinfo{publisher}{Association for Computational Linguistics}, \bibinfo{address}{Toronto, Canada}, \bibinfo{pages}{16190--16206}.
\newblock
\href{https://doi.org/10.18653/v1/2023.acl-long.895}{doi:\nolinkurl{10.18653/v1/2023.acl-long.895}}


\bibitem[Kotonya and Toni(2020)]%
        {kotonya-toni-2020-explainable-automated}
\bibfield{author}{\bibinfo{person}{Neema Kotonya} {and} \bibinfo{person}{Francesca Toni}.} \bibinfo{year}{2020}\natexlab{}.
\newblock \showarticletitle{Explainable Automated Fact-Checking for Public Health Claims}. In \bibinfo{booktitle}{\emph{Proceedings of the 2020 Conference on Empirical Methods in Natural Language Processing (EMNLP)}}, \bibfield{editor}{\bibinfo{person}{Bonnie Webber}, \bibinfo{person}{Trevor Cohn}, \bibinfo{person}{Yulan He}, {and} \bibinfo{person}{Yang Liu}} (Eds.). \bibinfo{publisher}{Association for Computational Linguistics}, \bibinfo{address}{Online}, \bibinfo{pages}{7740--7754}.
\newblock
\href{https://doi.org/10.18653/v1/2020.emnlp-main.623}{doi:\nolinkurl{10.18653/v1/2020.emnlp-main.623}}


\bibitem[Li et~al\mbox{.}(2021)]%
        {li2021paragraph}
\bibfield{author}{\bibinfo{person}{Xiangci Li}, \bibinfo{person}{Gully~A Burns}, {and} \bibinfo{person}{Nanyun Peng}.} \bibinfo{year}{2021}\natexlab{}.
\newblock \showarticletitle{A Paragraph-level Multi-task Learning Model for Scientific Fact-Verification.}. In \bibinfo{booktitle}{\emph{SDU@ AAAI}}.
\newblock


\bibitem[Li et~al\mbox{.}(2024)]%
        {li2024corpuslm}
\bibfield{author}{\bibinfo{person}{Xiaoxi Li}, \bibinfo{person}{Zhicheng Dou}, \bibinfo{person}{Yujia Zhou}, {and} \bibinfo{person}{Fangchao Liu}.} \bibinfo{year}{2024}\natexlab{}.
\newblock \showarticletitle{Corpuslm: Towards a unified language model on corpus for knowledge-intensive tasks}. In \bibinfo{booktitle}{\emph{Proceedings of the 47th International ACM SIGIR Conference on Research and Development in Information Retrieval}}. \bibinfo{pages}{26--37}.
\newblock


\bibitem[Lien et~al\mbox{.}(2019)]%
        {lien2019assumption}
\bibfield{author}{\bibinfo{person}{Yen-Chieh Lien}, \bibinfo{person}{Daniel Cohen}, {and} \bibinfo{person}{W~Bruce Croft}.} \bibinfo{year}{2019}\natexlab{}.
\newblock \showarticletitle{An assumption-free approach to the dynamic truncation of ranked lists}. In \bibinfo{booktitle}{\emph{Proceedings of the 2019 ACM SIGIR International Conference on Theory of Information Retrieval}}. \bibinfo{pages}{79--82}.
\newblock


\bibitem[Lindberg et~al\mbox{.}(1993)]%
        {lindberg1993unified}
\bibfield{author}{\bibinfo{person}{Donald~AB Lindberg}, \bibinfo{person}{Betsy~L Humphreys}, {and} \bibinfo{person}{Alexa~T McCray}.} \bibinfo{year}{1993}\natexlab{}.
\newblock \showarticletitle{The unified medical language system}.
\newblock \bibinfo{journal}{\emph{Yearbook of medical informatics}} \bibinfo{volume}{2}, \bibinfo{number}{01} (\bibinfo{year}{1993}), \bibinfo{pages}{41--51}.
\newblock


\bibitem[Lipscomb(2000)]%
        {lipscomb2000medical}
\bibfield{author}{\bibinfo{person}{Carolyn~E Lipscomb}.} \bibinfo{year}{2000}\natexlab{}.
\newblock \showarticletitle{Medical subject headings (MeSH)}.
\newblock \bibinfo{journal}{\emph{Bulletin of the Medical Library Association}} \bibinfo{volume}{88}, \bibinfo{number}{3} (\bibinfo{year}{2000}), \bibinfo{pages}{265}.
\newblock


\bibitem[Liu et~al\mbox{.}(2025)]%
        {10.1145/3696410.3714554}
\bibfield{author}{\bibinfo{person}{Qi Liu}, \bibinfo{person}{Bo Wang}, \bibinfo{person}{Nan Wang}, {and} \bibinfo{person}{Jiaxin Mao}.} \bibinfo{year}{2025}\natexlab{}.
\newblock \showarticletitle{Leveraging Passage Embeddings for Efficient Listwise Reranking with Large Language Models}. In \bibinfo{booktitle}{\emph{Proceedings of the ACM on Web Conference 2025}} (Sydney NSW, Australia) \emph{(\bibinfo{series}{WWW '25})}. \bibinfo{publisher}{Association for Computing Machinery}, \bibinfo{address}{New York, NY, USA}, \bibinfo{pages}{4274–4283}.
\newblock
\showISBNx{9798400712746}
\href{https://doi.org/10.1145/3696410.3714554}{doi:\nolinkurl{10.1145/3696410.3714554}}


\bibitem[Lo et~al\mbox{.}(2024)]%
        {10.1145/3659096}
\bibfield{author}{\bibinfo{person}{Kyle Lo}, \bibinfo{person}{Joseph~Chee Chang}, \bibinfo{person}{Andrew Head}, \bibinfo{person}{Jonathan Bragg}, \bibinfo{person}{Amy~X. Zhang}, \bibinfo{person}{Cassidy Trier}, \bibinfo{person}{Chloe Anastasiades}, \bibinfo{person}{Tal August}, \bibinfo{person}{Russell Authur}, \bibinfo{person}{Danielle Bragg}, \bibinfo{person}{Erin Bransom}, \bibinfo{person}{Isabel Cachola}, \bibinfo{person}{Stefan Candra}, \bibinfo{person}{Yoganand Chandrasekhar}, \bibinfo{person}{Yen-Sung Chen}, \bibinfo{person}{Evie Yu-Yen Cheng}, \bibinfo{person}{Yvonne Chou}, \bibinfo{person}{Doug Downey}, \bibinfo{person}{Rob Evans}, \bibinfo{person}{Raymond Fok}, \bibinfo{person}{Fangzhou Hu}, \bibinfo{person}{Regan Huff}, \bibinfo{person}{Dongyeop Kang}, \bibinfo{person}{Tae~Soo Kim}, \bibinfo{person}{Rodney Kinney}, \bibinfo{person}{Aniket Kittur}, \bibinfo{person}{Hyeonsu~B. Kang}, \bibinfo{person}{Egor Klevak}, \bibinfo{person}{Bailey Kuehl}, \bibinfo{person}{Michael~J. Langan},
  \bibinfo{person}{Matt Latzke}, \bibinfo{person}{Jaron Lochner}, \bibinfo{person}{Kelsey MacMillan}, \bibinfo{person}{Eric Marsh}, \bibinfo{person}{Tyler Murray}, \bibinfo{person}{Aakanksha Naik}, \bibinfo{person}{Ngoc-Uyen Nguyen}, \bibinfo{person}{Srishti Palani}, \bibinfo{person}{Soya Park}, \bibinfo{person}{Caroline Paulic}, \bibinfo{person}{Napol Rachatasumrit}, \bibinfo{person}{Smita Rao}, \bibinfo{person}{Paul Sayre}, \bibinfo{person}{Zejiang Shen}, \bibinfo{person}{Pao Siangliulue}, \bibinfo{person}{Luca Soldaini}, \bibinfo{person}{Huy Tran}, \bibinfo{person}{Madeleine van Zuylen}, \bibinfo{person}{Lucy~Lu Wang}, \bibinfo{person}{Christopher Wilhelm}, \bibinfo{person}{Caroline Wu}, \bibinfo{person}{Jiangjiang Yang}, \bibinfo{person}{Angele Zamarron}, \bibinfo{person}{Marti~A. Hearst}, {and} \bibinfo{person}{Daniel~S. Weld}.} \bibinfo{year}{2024}\natexlab{}.
\newblock \showarticletitle{The Semantic Reader Project}.
\newblock \bibinfo{journal}{\emph{Commun. ACM}} \bibinfo{volume}{67}, \bibinfo{number}{10} (\bibinfo{date}{Sept.} \bibinfo{year}{2024}), \bibinfo{pages}{50–61}.
\newblock
\showISSN{0001-0782}
\href{https://doi.org/10.1145/3659096}{doi:\nolinkurl{10.1145/3659096}}


\bibitem[Lu et~al\mbox{.}(2023)]%
        {lu-etal-2023-scitab}
\bibfield{author}{\bibinfo{person}{Xinyuan Lu}, \bibinfo{person}{Liangming Pan}, \bibinfo{person}{Qian Liu}, \bibinfo{person}{Preslav Nakov}, {and} \bibinfo{person}{Min-Yen Kan}.} \bibinfo{year}{2023}\natexlab{}.
\newblock \showarticletitle{{SCITAB}: A Challenging Benchmark for Compositional Reasoning and Claim Verification on Scientific Tables}. In \bibinfo{booktitle}{\emph{Proceedings of the 2023 Conference on Empirical Methods in Natural Language Processing}}, \bibfield{editor}{\bibinfo{person}{Houda Bouamor}, \bibinfo{person}{Juan Pino}, {and} \bibinfo{person}{Kalika Bali}} (Eds.). \bibinfo{publisher}{Association for Computational Linguistics}, \bibinfo{address}{Singapore}, \bibinfo{pages}{7787--7813}.
\newblock
\href{https://doi.org/10.18653/v1/2023.emnlp-main.483}{doi:\nolinkurl{10.18653/v1/2023.emnlp-main.483}}


\bibitem[Luengo and Garc{\'\i}a-Mar{\'\i}n(2020)]%
        {luengo2020performance}
\bibfield{author}{\bibinfo{person}{Mar{\'\i}a Luengo} {and} \bibinfo{person}{David Garc{\'\i}a-Mar{\'\i}n}.} \bibinfo{year}{2020}\natexlab{}.
\newblock \showarticletitle{The performance of truth: politicians, fact-checking journalism, and the struggle to tackle COVID-19 misinformation}.
\newblock \bibinfo{journal}{\emph{American Journal of Cultural Sociology}} \bibinfo{volume}{8}, \bibinfo{number}{3} (\bibinfo{year}{2020}), \bibinfo{pages}{405}.
\newblock


\bibitem[Ma et~al\mbox{.}(2022)]%
        {ma2022incorporating}
\bibfield{author}{\bibinfo{person}{Yixiao Ma}, \bibinfo{person}{Qingyao Ai}, \bibinfo{person}{Yueyue Wu}, \bibinfo{person}{Yunqiu Shao}, \bibinfo{person}{Yiqun Liu}, \bibinfo{person}{Min Zhang}, {and} \bibinfo{person}{Shaoping Ma}.} \bibinfo{year}{2022}\natexlab{}.
\newblock \showarticletitle{Incorporating retrieval information into the truncation of ranking lists for better legal search}. In \bibinfo{booktitle}{\emph{Proceedings of the 45th International ACM SIGIR Conference on Research and Development in Information Retrieval}}. \bibinfo{pages}{438--448}.
\newblock


\bibitem[Ma et~al\mbox{.}(2024)]%
        {ma2024mmlongbenchdocbenchmarkinglongcontextdocument}
\bibfield{author}{\bibinfo{person}{Yubo Ma}, \bibinfo{person}{Yuhang Zang}, \bibinfo{person}{Liangyu Chen}, \bibinfo{person}{Meiqi Chen}, \bibinfo{person}{Yizhu Jiao}, \bibinfo{person}{Xinze Li}, \bibinfo{person}{Xinyuan Lu}, \bibinfo{person}{Ziyu Liu}, \bibinfo{person}{Yan Ma}, \bibinfo{person}{Xiaoyi Dong}, \bibinfo{person}{Pan Zhang}, \bibinfo{person}{Liangming Pan}, \bibinfo{person}{Yu-Gang Jiang}, \bibinfo{person}{Jiaqi Wang}, \bibinfo{person}{Yixin Cao}, {and} \bibinfo{person}{Aixin Sun}.} \bibinfo{year}{2024}\natexlab{}.
\newblock \bibinfo{title}{MMLongBench-Doc: Benchmarking Long-context Document Understanding with Visualizations}.
\newblock
\showeprint[arxiv]{2407.01523}~[cs.CV]
\urldef\tempurl%
\url{https://arxiv.org/abs/2407.01523}
\showURL{%
\tempurl}


\bibitem[Meng et~al\mbox{.}(2024)]%
        {meng2024ranked}
\bibfield{author}{\bibinfo{person}{Chuan Meng}, \bibinfo{person}{Negar Arabzadeh}, \bibinfo{person}{Arian Askari}, \bibinfo{person}{Mohammad Aliannejadi}, {and} \bibinfo{person}{Maarten de Rijke}.} \bibinfo{year}{2024}\natexlab{}.
\newblock \showarticletitle{Ranked list truncation for large language model-based re-ranking}. In \bibinfo{booktitle}{\emph{Proceedings of the 47th International ACM SIGIR Conference on Research and Development in Information Retrieval}}. \bibinfo{pages}{141--151}.
\newblock


\bibitem[Mohr et~al\mbox{.}(2022)]%
        {mohr-etal-2022-covert}
\bibfield{author}{\bibinfo{person}{Isabelle Mohr}, \bibinfo{person}{Amelie W{\"u}hrl}, {and} \bibinfo{person}{Roman Klinger}.} \bibinfo{year}{2022}\natexlab{}.
\newblock \showarticletitle{{C}o{VERT}: A Corpus of Fact-checked Biomedical {COVID}-19 Tweets}. In \bibinfo{booktitle}{\emph{Proceedings of the Thirteenth Language Resources and Evaluation Conference}}, \bibfield{editor}{\bibinfo{person}{Nicoletta Calzolari}, \bibinfo{person}{Fr{\'e}d{\'e}ric B{\'e}chet}, \bibinfo{person}{Philippe Blache}, \bibinfo{person}{Khalid Choukri}, \bibinfo{person}{Christopher Cieri}, \bibinfo{person}{Thierry Declerck}, \bibinfo{person}{Sara Goggi}, \bibinfo{person}{Hitoshi Isahara}, \bibinfo{person}{Bente Maegaard}, \bibinfo{person}{Joseph Mariani}, \bibinfo{person}{H{\'e}l{\`e}ne Mazo}, \bibinfo{person}{Jan Odijk}, {and} \bibinfo{person}{Stelios Piperidis}} (Eds.). \bibinfo{publisher}{European Language Resources Association}, \bibinfo{address}{Marseille, France}, \bibinfo{pages}{244--257}.
\newblock
\urldef\tempurl%
\url{https://aclanthology.org/2022.lrec-1.26}
\showURL{%
\tempurl}


\bibitem[Momii et~al\mbox{.}(2024)]%
        {momii-etal-2024-rag}
\bibfield{author}{\bibinfo{person}{Yuki Momii}, \bibinfo{person}{Tetsuya Takiguchi}, {and} \bibinfo{person}{Yasuo Ariki}.} \bibinfo{year}{2024}\natexlab{}.
\newblock \showarticletitle{{RAG}-Fusion Based Information Retrieval for Fact-Checking}. In \bibinfo{booktitle}{\emph{Proceedings of the Seventh Fact Extraction and VERification Workshop (FEVER)}}, \bibfield{editor}{\bibinfo{person}{Michael Schlichtkrull}, \bibinfo{person}{Yulong Chen}, \bibinfo{person}{Chenxi Whitehouse}, \bibinfo{person}{Zhenyun Deng}, \bibinfo{person}{Mubashara Akhtar}, \bibinfo{person}{Rami Aly}, \bibinfo{person}{Zhijiang Guo}, \bibinfo{person}{Christos Christodoulopoulos}, \bibinfo{person}{Oana Cocarascu}, \bibinfo{person}{Arpit Mittal}, \bibinfo{person}{James Thorne}, {and} \bibinfo{person}{Andreas Vlachos}} (Eds.). \bibinfo{publisher}{Association for Computational Linguistics}, \bibinfo{address}{Miami, Florida, USA}, \bibinfo{pages}{47--54}.
\newblock
\href{https://doi.org/10.18653/v1/2024.fever-1.4}{doi:\nolinkurl{10.18653/v1/2024.fever-1.4}}


\bibitem[Nogueira et~al\mbox{.}(2020)]%
        {nogueira-etal-2020-document}
\bibfield{author}{\bibinfo{person}{Rodrigo Nogueira}, \bibinfo{person}{Zhiying Jiang}, \bibinfo{person}{Ronak Pradeep}, {and} \bibinfo{person}{Jimmy Lin}.} \bibinfo{year}{2020}\natexlab{}.
\newblock \showarticletitle{Document Ranking with a Pretrained Sequence-to-Sequence Model}. In \bibinfo{booktitle}{\emph{Findings of the Association for Computational Linguistics: EMNLP 2020}}, \bibfield{editor}{\bibinfo{person}{Trevor Cohn}, \bibinfo{person}{Yulan He}, {and} \bibinfo{person}{Yang Liu}} (Eds.). \bibinfo{publisher}{Association for Computational Linguistics}, \bibinfo{address}{Online}, \bibinfo{pages}{708--718}.
\newblock
\href{https://doi.org/10.18653/v1/2020.findings-emnlp.63}{doi:\nolinkurl{10.18653/v1/2020.findings-emnlp.63}}


\bibitem[Papadopoulos et~al\mbox{.}(2023)]%
        {papadopoulos2023synthetic}
\bibfield{author}{\bibinfo{person}{Stefanos-Iordanis Papadopoulos}, \bibinfo{person}{Christos Koutlis}, \bibinfo{person}{Symeon Papadopoulos}, {and} \bibinfo{person}{Panagiotis Petrantonakis}.} \bibinfo{year}{2023}\natexlab{}.
\newblock \showarticletitle{Synthetic misinformers: Generating and combating multimodal misinformation}. In \bibinfo{booktitle}{\emph{Proceedings of the 2nd ACM International Workshop on Multimedia AI against Disinformation}}. \bibinfo{pages}{36--44}.
\newblock


\bibitem[Papadopoulos et~al\mbox{.}(2024)]%
        {papadopoulos2024verite}
\bibfield{author}{\bibinfo{person}{Stefanos-Iordanis Papadopoulos}, \bibinfo{person}{Christos Koutlis}, \bibinfo{person}{Symeon Papadopoulos}, {and} \bibinfo{person}{Panagiotis~C Petrantonakis}.} \bibinfo{year}{2024}\natexlab{}.
\newblock \showarticletitle{VERITE: a Robust benchmark for multimodal misinformation detection accounting for unimodal bias}.
\newblock \bibinfo{journal}{\emph{International Journal of Multimedia Information Retrieval}} \bibinfo{volume}{13}, \bibinfo{number}{1} (\bibinfo{year}{2024}), \bibinfo{pages}{4}.
\newblock


\bibitem[Parry et~al\mbox{.}(2024)]%
        {parry2024context}
\bibfield{author}{\bibinfo{person}{Andrew Parry}, \bibinfo{person}{Debasis Ganguly}, {and} \bibinfo{person}{Manish Chandra}.} \bibinfo{year}{2024}\natexlab{}.
\newblock \showarticletitle{In-Context Learning" or: How I learned to stop worrying and love" Applied Information Retrieval}. In \bibinfo{booktitle}{\emph{Proceedings of the 47th International ACM SIGIR Conference on Research and Development in Information Retrieval}}. \bibinfo{pages}{14--25}.
\newblock


\bibitem[Pradeep et~al\mbox{.}(2021)]%
        {pradeep-etal-2021-scientific}
\bibfield{author}{\bibinfo{person}{Ronak Pradeep}, \bibinfo{person}{Xueguang Ma}, \bibinfo{person}{Rodrigo Nogueira}, {and} \bibinfo{person}{Jimmy Lin}.} \bibinfo{year}{2021}\natexlab{}.
\newblock \showarticletitle{Scientific Claim Verification with {V}er{T}5erini}. In \bibinfo{booktitle}{\emph{Proceedings of the 12th International Workshop on Health Text Mining and Information Analysis}}, \bibfield{editor}{\bibinfo{person}{Eben Holderness}, \bibinfo{person}{Antonio Jimeno~Yepes}, \bibinfo{person}{Alberto Lavelli}, \bibinfo{person}{Anne-Lyse Minard}, \bibinfo{person}{James Pustejovsky}, {and} \bibinfo{person}{Fabio Rinaldi}} (Eds.). \bibinfo{publisher}{Association for Computational Linguistics}, \bibinfo{address}{online}, \bibinfo{pages}{94--103}.
\newblock
\urldef\tempurl%
\url{https://aclanthology.org/2021.louhi-1.11/}
\showURL{%
\tempurl}


\bibitem[Qi et~al\mbox{.}(2024)]%
        {qi2024sniffer}
\bibfield{author}{\bibinfo{person}{Peng Qi}, \bibinfo{person}{Zehong Yan}, \bibinfo{person}{Wynne Hsu}, {and} \bibinfo{person}{Mong~Li Lee}.} \bibinfo{year}{2024}\natexlab{}.
\newblock \showarticletitle{SNIFFER: Multimodal Large Language Model for Explainable Out-of-Context Misinformation Detection}. In \bibinfo{booktitle}{\emph{Proceedings of the IEEE/CVF Conference on Computer Vision and Pattern Recognition}}. \bibinfo{pages}{13052--13062}.
\newblock


\bibitem[Raina and Gales(2024)]%
        {raina-gales-2024-question}
\bibfield{author}{\bibinfo{person}{Vatsal Raina} {and} \bibinfo{person}{Mark Gales}.} \bibinfo{year}{2024}\natexlab{}.
\newblock \showarticletitle{Question-Based Retrieval using Atomic Units for Enterprise {RAG}}. In \bibinfo{booktitle}{\emph{Proceedings of the Seventh Fact Extraction and VERification Workshop (FEVER)}}, \bibfield{editor}{\bibinfo{person}{Michael Schlichtkrull}, \bibinfo{person}{Yulong Chen}, \bibinfo{person}{Chenxi Whitehouse}, \bibinfo{person}{Zhenyun Deng}, \bibinfo{person}{Mubashara Akhtar}, \bibinfo{person}{Rami Aly}, \bibinfo{person}{Zhijiang Guo}, \bibinfo{person}{Christos Christodoulopoulos}, \bibinfo{person}{Oana Cocarascu}, \bibinfo{person}{Arpit Mittal}, \bibinfo{person}{James Thorne}, {and} \bibinfo{person}{Andreas Vlachos}} (Eds.). \bibinfo{publisher}{Association for Computational Linguistics}, \bibinfo{address}{Miami, Florida, USA}, \bibinfo{pages}{219--233}.
\newblock
\href{https://doi.org/10.18653/v1/2024.fever-1.25}{doi:\nolinkurl{10.18653/v1/2024.fever-1.25}}


\bibitem[Roberts et~al\mbox{.}(2024)]%
        {roberts2024scifibench}
\bibfield{author}{\bibinfo{person}{Jonathan Roberts}, \bibinfo{person}{Kai Han}, \bibinfo{person}{Neil Houlsby}, {and} \bibinfo{person}{Samuel Albanie}.} \bibinfo{year}{2024}\natexlab{}.
\newblock \showarticletitle{SciFIBench: Benchmarking Large Multimodal Models for Scientific Figure Interpretation}.
\newblock \bibinfo{journal}{\emph{arXiv preprint arXiv:2405.08807}} (\bibinfo{year}{2024}).
\newblock


\bibitem[Saakyan et~al\mbox{.}(2021)]%
        {saakyan-etal-2021-covid}
\bibfield{author}{\bibinfo{person}{Arkadiy Saakyan}, \bibinfo{person}{Tuhin Chakrabarty}, {and} \bibinfo{person}{Smaranda Muresan}.} \bibinfo{year}{2021}\natexlab{}.
\newblock \showarticletitle{{COVID}-Fact: Fact Extraction and Verification of Real-World Claims on {COVID}-19 Pandemic}. In \bibinfo{booktitle}{\emph{Proceedings of the 59th Annual Meeting of the Association for Computational Linguistics and the 11th International Joint Conference on Natural Language Processing (Volume 1: Long Papers)}}, \bibfield{editor}{\bibinfo{person}{Chengqing Zong}, \bibinfo{person}{Fei Xia}, \bibinfo{person}{Wenjie Li}, {and} \bibinfo{person}{Roberto Navigli}} (Eds.). \bibinfo{publisher}{Association for Computational Linguistics}, \bibinfo{address}{Online}, \bibinfo{pages}{2116--2129}.
\newblock
\href{https://doi.org/10.18653/v1/2021.acl-long.165}{doi:\nolinkurl{10.18653/v1/2021.acl-long.165}}


\bibitem[Salemi and Zamani(2024a)]%
        {salemi2024evaluating}
\bibfield{author}{\bibinfo{person}{Alireza Salemi} {and} \bibinfo{person}{Hamed Zamani}.} \bibinfo{year}{2024}\natexlab{a}.
\newblock \showarticletitle{Evaluating retrieval quality in retrieval-augmented generation}. In \bibinfo{booktitle}{\emph{Proceedings of the 47th International ACM SIGIR Conference on Research and Development in Information Retrieval}}. \bibinfo{pages}{2395--2400}.
\newblock


\bibitem[Salemi and Zamani(2024b)]%
        {salemi2024towards}
\bibfield{author}{\bibinfo{person}{Alireza Salemi} {and} \bibinfo{person}{Hamed Zamani}.} \bibinfo{year}{2024}\natexlab{b}.
\newblock \showarticletitle{Towards a search engine for machines: Unified ranking for multiple retrieval-augmented large language models}. In \bibinfo{booktitle}{\emph{Proceedings of the 47th International ACM SIGIR Conference on Research and Development in Information Retrieval}}. \bibinfo{pages}{741--751}.
\newblock


\bibitem[Sarrouti et~al\mbox{.}(2021)]%
        {sarrouti-etal-2021-evidence-based}
\bibfield{author}{\bibinfo{person}{Mourad Sarrouti}, \bibinfo{person}{Asma Ben~Abacha}, \bibinfo{person}{Yassine Mrabet}, {and} \bibinfo{person}{Dina Demner-Fushman}.} \bibinfo{year}{2021}\natexlab{}.
\newblock \showarticletitle{Evidence-based Fact-Checking of Health-related Claims}. In \bibinfo{booktitle}{\emph{Findings of the Association for Computational Linguistics: EMNLP 2021}}, \bibfield{editor}{\bibinfo{person}{Marie-Francine Moens}, \bibinfo{person}{Xuanjing Huang}, \bibinfo{person}{Lucia Specia}, {and} \bibinfo{person}{Scott Wen-tau Yih}} (Eds.). \bibinfo{publisher}{Association for Computational Linguistics}, \bibinfo{address}{Punta Cana, Dominican Republic}, \bibinfo{pages}{3499--3512}.
\newblock
\href{https://doi.org/10.18653/v1/2021.findings-emnlp.297}{doi:\nolinkurl{10.18653/v1/2021.findings-emnlp.297}}


\bibitem[Sauchuk et~al\mbox{.}(2022)]%
        {sauchuk2022role}
\bibfield{author}{\bibinfo{person}{Artsiom Sauchuk}, \bibinfo{person}{James Thorne}, \bibinfo{person}{Alon Halevy}, \bibinfo{person}{Nicola Tonellotto}, {and} \bibinfo{person}{Fabrizio Silvestri}.} \bibinfo{year}{2022}\natexlab{}.
\newblock \showarticletitle{On the role of relevance in natural language processing tasks}. In \bibinfo{booktitle}{\emph{Proceedings of the 45th International ACM SIGIR Conference on Research and Development in Information Retrieval}}. \bibinfo{pages}{1785--1789}.
\newblock


\bibitem[Schlichtkrull et~al\mbox{.}(2024a)]%
        {schlichtkrull-etal-2024-automated}
\bibfield{author}{\bibinfo{person}{Michael Schlichtkrull}, \bibinfo{person}{Yulong Chen}, \bibinfo{person}{Chenxi Whitehouse}, \bibinfo{person}{Zhenyun Deng}, \bibinfo{person}{Mubashara Akhtar}, \bibinfo{person}{Rami Aly}, \bibinfo{person}{Zhijiang Guo}, \bibinfo{person}{Christos Christodoulopoulos}, \bibinfo{person}{Oana Cocarascu}, \bibinfo{person}{Arpit Mittal}, \bibinfo{person}{James Thorne}, {and} \bibinfo{person}{Andreas Vlachos}.} \bibinfo{year}{2024}\natexlab{a}.
\newblock \showarticletitle{The Automated Verification of Textual Claims ({AV}eri{T}e{C}) Shared Task}. In \bibinfo{booktitle}{\emph{Proceedings of the Seventh Fact Extraction and VERification Workshop (FEVER)}}, \bibfield{editor}{\bibinfo{person}{Michael Schlichtkrull}, \bibinfo{person}{Yulong Chen}, \bibinfo{person}{Chenxi Whitehouse}, \bibinfo{person}{Zhenyun Deng}, \bibinfo{person}{Mubashara Akhtar}, \bibinfo{person}{Rami Aly}, \bibinfo{person}{Zhijiang Guo}, \bibinfo{person}{Christos Christodoulopoulos}, \bibinfo{person}{Oana Cocarascu}, \bibinfo{person}{Arpit Mittal}, \bibinfo{person}{James Thorne}, {and} \bibinfo{person}{Andreas Vlachos}} (Eds.). \bibinfo{publisher}{Association for Computational Linguistics}, \bibinfo{address}{Miami, Florida, USA}, \bibinfo{pages}{1--26}.
\newblock
\href{https://doi.org/10.18653/v1/2024.fever-1.1}{doi:\nolinkurl{10.18653/v1/2024.fever-1.1}}


\bibitem[Schlichtkrull et~al\mbox{.}(2024b)]%
        {schlichtkrull2024averitec}
\bibfield{author}{\bibinfo{person}{Michael Schlichtkrull}, \bibinfo{person}{Zhijiang Guo}, {and} \bibinfo{person}{Andreas Vlachos}.} \bibinfo{year}{2024}\natexlab{b}.
\newblock \showarticletitle{Averitec: A dataset for real-world claim verification with evidence from the web}.
\newblock \bibinfo{journal}{\emph{Advances in Neural Information Processing Systems}}  \bibinfo{volume}{36} (\bibinfo{year}{2024}).
\newblock


\bibitem[Schlichtkrull et~al\mbox{.}(2021)]%
        {schlichtkrull-etal-2021-joint}
\bibfield{author}{\bibinfo{person}{Michael~Sejr Schlichtkrull}, \bibinfo{person}{Vladimir Karpukhin}, \bibinfo{person}{Barlas Oguz}, \bibinfo{person}{Mike Lewis}, \bibinfo{person}{Wen-tau Yih}, {and} \bibinfo{person}{Sebastian Riedel}.} \bibinfo{year}{2021}\natexlab{}.
\newblock \showarticletitle{Joint Verification and Reranking for Open Fact Checking Over Tables}. In \bibinfo{booktitle}{\emph{Proceedings of the 59th Annual Meeting of the Association for Computational Linguistics and the 11th International Joint Conference on Natural Language Processing (Volume 1: Long Papers)}}, \bibfield{editor}{\bibinfo{person}{Chengqing Zong}, \bibinfo{person}{Fei Xia}, \bibinfo{person}{Wenjie Li}, {and} \bibinfo{person}{Roberto Navigli}} (Eds.). \bibinfo{publisher}{Association for Computational Linguistics}, \bibinfo{address}{Online}, \bibinfo{pages}{6787--6799}.
\newblock
\href{https://doi.org/10.18653/v1/2021.acl-long.529}{doi:\nolinkurl{10.18653/v1/2021.acl-long.529}}


\bibitem[Sevgili et~al\mbox{.}(2024)]%
        {sevgili-etal-2024-uhh}
\bibfield{author}{\bibinfo{person}{{\"O}zge Sevgili}, \bibinfo{person}{Irina Nikishina}, \bibinfo{person}{Seid~Muhie Yimam}, \bibinfo{person}{Martin Semmann}, {and} \bibinfo{person}{Chris Biemann}.} \bibinfo{year}{2024}\natexlab{}.
\newblock \showarticletitle{{UHH} at {AV}eri{T}e{C}: {RAG} for Fact-Checking with Real-World Claims}. In \bibinfo{booktitle}{\emph{Proceedings of the Seventh Fact Extraction and VERification Workshop (FEVER)}}, \bibfield{editor}{\bibinfo{person}{Michael Schlichtkrull}, \bibinfo{person}{Yulong Chen}, \bibinfo{person}{Chenxi Whitehouse}, \bibinfo{person}{Zhenyun Deng}, \bibinfo{person}{Mubashara Akhtar}, \bibinfo{person}{Rami Aly}, \bibinfo{person}{Zhijiang Guo}, \bibinfo{person}{Christos Christodoulopoulos}, \bibinfo{person}{Oana Cocarascu}, \bibinfo{person}{Arpit Mittal}, \bibinfo{person}{James Thorne}, {and} \bibinfo{person}{Andreas Vlachos}} (Eds.). \bibinfo{publisher}{Association for Computational Linguistics}, \bibinfo{address}{Miami, Florida, USA}, \bibinfo{pages}{55--63}.
\newblock
\href{https://doi.org/10.18653/v1/2024.fever-1.5}{doi:\nolinkurl{10.18653/v1/2024.fever-1.5}}


\bibitem[Shen et~al\mbox{.}(2022)]%
        {shen-etal-2022-vila}
\bibfield{author}{\bibinfo{person}{Zejiang Shen}, \bibinfo{person}{Kyle Lo}, \bibinfo{person}{Lucy~Lu Wang}, \bibinfo{person}{Bailey Kuehl}, \bibinfo{person}{Daniel~S. Weld}, {and} \bibinfo{person}{Doug Downey}.} \bibinfo{year}{2022}\natexlab{}.
\newblock \showarticletitle{{VILA}: Improving Structured Content Extraction from Scientific {PDF}s Using Visual Layout Groups}.
\newblock \bibinfo{journal}{\emph{Transactions of the Association for Computational Linguistics}}  \bibinfo{volume}{10} (\bibinfo{year}{2022}), \bibinfo{pages}{376--392}.
\newblock
\href{https://doi.org/10.1162/tacl_a_00466}{doi:\nolinkurl{10.1162/tacl_a_00466}}


\bibitem[Shi et~al\mbox{.}(2023)]%
        {shi2023large}
\bibfield{author}{\bibinfo{person}{Freda Shi}, \bibinfo{person}{Xinyun Chen}, \bibinfo{person}{Kanishka Misra}, \bibinfo{person}{Nathan Scales}, \bibinfo{person}{David Dohan}, \bibinfo{person}{Ed~H Chi}, \bibinfo{person}{Nathanael Sch{\"a}rli}, {and} \bibinfo{person}{Denny Zhou}.} \bibinfo{year}{2023}\natexlab{}.
\newblock \showarticletitle{Large language models can be easily distracted by irrelevant context}. In \bibinfo{booktitle}{\emph{International Conference on Machine Learning}}. PMLR, \bibinfo{pages}{31210--31227}.
\newblock


\bibitem[Shi et~al\mbox{.}(2021)]%
        {shi-etal-2021-logic}
\bibfield{author}{\bibinfo{person}{Qi Shi}, \bibinfo{person}{Yu Zhang}, \bibinfo{person}{Qingyu Yin}, {and} \bibinfo{person}{Ting Liu}.} \bibinfo{year}{2021}\natexlab{}.
\newblock \showarticletitle{Logic-level Evidence Retrieval and Graph-based Verification Network for Table-based Fact Verification}. In \bibinfo{booktitle}{\emph{Proceedings of the 2021 Conference on Empirical Methods in Natural Language Processing}}, \bibfield{editor}{\bibinfo{person}{Marie-Francine Moens}, \bibinfo{person}{Xuanjing Huang}, \bibinfo{person}{Lucia Specia}, {and} \bibinfo{person}{Scott Wen-tau Yih}} (Eds.). \bibinfo{publisher}{Association for Computational Linguistics}, \bibinfo{address}{Online and Punta Cana, Dominican Republic}, \bibinfo{pages}{175--184}.
\newblock
\href{https://doi.org/10.18653/v1/2021.emnlp-main.16}{doi:\nolinkurl{10.18653/v1/2021.emnlp-main.16}}


\bibitem[Singal et~al\mbox{.}(2024)]%
        {singal-etal-2024-evidence}
\bibfield{author}{\bibinfo{person}{Ronit Singal}, \bibinfo{person}{Pransh Patwa}, \bibinfo{person}{Parth Patwa}, \bibinfo{person}{Aman Chadha}, {and} \bibinfo{person}{Amitava Das}.} \bibinfo{year}{2024}\natexlab{}.
\newblock \showarticletitle{Evidence-backed Fact Checking using {RAG} and Few-Shot In-Context Learning with {LLM}s}. In \bibinfo{booktitle}{\emph{Proceedings of the Seventh Fact Extraction and VERification Workshop (FEVER)}}, \bibfield{editor}{\bibinfo{person}{Michael Schlichtkrull}, \bibinfo{person}{Yulong Chen}, \bibinfo{person}{Chenxi Whitehouse}, \bibinfo{person}{Zhenyun Deng}, \bibinfo{person}{Mubashara Akhtar}, \bibinfo{person}{Rami Aly}, \bibinfo{person}{Zhijiang Guo}, \bibinfo{person}{Christos Christodoulopoulos}, \bibinfo{person}{Oana Cocarascu}, \bibinfo{person}{Arpit Mittal}, \bibinfo{person}{James Thorne}, {and} \bibinfo{person}{Andreas Vlachos}} (Eds.). \bibinfo{publisher}{Association for Computational Linguistics}, \bibinfo{address}{Miami, Florida, USA}, \bibinfo{pages}{91--98}.
\newblock
\href{https://doi.org/10.18653/v1/2024.fever-1.10}{doi:\nolinkurl{10.18653/v1/2024.fever-1.10}}


\bibitem[Stevenson and Bin-Hezam(2023)]%
        {stevenson2023stopping}
\bibfield{author}{\bibinfo{person}{Mark Stevenson} {and} \bibinfo{person}{Reem Bin-Hezam}.} \bibinfo{year}{2023}\natexlab{}.
\newblock \showarticletitle{Stopping Methods for Technology-assisted Reviews Based on Point Processes}.
\newblock \bibinfo{journal}{\emph{ACM Transactions on Information Systems}} \bibinfo{volume}{42}, \bibinfo{number}{3} (\bibinfo{year}{2023}), \bibinfo{pages}{1--37}.
\newblock


\bibitem[Sun et~al\mbox{.}(2023)]%
        {sun-etal-2023-chatgpt}
\bibfield{author}{\bibinfo{person}{Weiwei Sun}, \bibinfo{person}{Lingyong Yan}, \bibinfo{person}{Xinyu Ma}, \bibinfo{person}{Shuaiqiang Wang}, \bibinfo{person}{Pengjie Ren}, \bibinfo{person}{Zhumin Chen}, \bibinfo{person}{Dawei Yin}, {and} \bibinfo{person}{Zhaochun Ren}.} \bibinfo{year}{2023}\natexlab{}.
\newblock \showarticletitle{Is {C}hat{GPT} Good at Search? Investigating Large Language Models as Re-Ranking Agents}. In \bibinfo{booktitle}{\emph{Proceedings of the 2023 Conference on Empirical Methods in Natural Language Processing}}, \bibfield{editor}{\bibinfo{person}{Houda Bouamor}, \bibinfo{person}{Juan Pino}, {and} \bibinfo{person}{Kalika Bali}} (Eds.). \bibinfo{publisher}{Association for Computational Linguistics}, \bibinfo{address}{Singapore}, \bibinfo{pages}{14918--14937}.
\newblock
\href{https://doi.org/10.18653/v1/2023.emnlp-main.923}{doi:\nolinkurl{10.18653/v1/2023.emnlp-main.923}}


\bibitem[Tanaka et~al\mbox{.}(2023)]%
        {tanaka2023slidevqa}
\bibfield{author}{\bibinfo{person}{Ryota Tanaka}, \bibinfo{person}{Kyosuke Nishida}, \bibinfo{person}{Kosuke Nishida}, \bibinfo{person}{Taku Hasegawa}, \bibinfo{person}{Itsumi Saito}, {and} \bibinfo{person}{Kuniko Saito}.} \bibinfo{year}{2023}\natexlab{}.
\newblock \showarticletitle{Slidevqa: A dataset for document visual question answering on multiple images}. In \bibinfo{booktitle}{\emph{Proceedings of the AAAI Conference on Artificial Intelligence}}, Vol.~\bibinfo{volume}{37}. \bibinfo{pages}{13636--13645}.
\newblock


\bibitem[Tang et~al\mbox{.}(2024)]%
        {tang-etal-2024-minicheck}
\bibfield{author}{\bibinfo{person}{Liyan Tang}, \bibinfo{person}{Philippe Laban}, {and} \bibinfo{person}{Greg Durrett}.} \bibinfo{year}{2024}\natexlab{}.
\newblock \showarticletitle{{M}ini{C}heck: Efficient Fact-Checking of {LLM}s on Grounding Documents}. In \bibinfo{booktitle}{\emph{Proceedings of the 2024 Conference on Empirical Methods in Natural Language Processing}}, \bibfield{editor}{\bibinfo{person}{Yaser Al-Onaizan}, \bibinfo{person}{Mohit Bansal}, {and} \bibinfo{person}{Yun-Nung Chen}} (Eds.). \bibinfo{publisher}{Association for Computational Linguistics}, \bibinfo{address}{Miami, Florida, USA}, \bibinfo{pages}{8818--8847}.
\newblock
\href{https://doi.org/10.18653/v1/2024.emnlp-main.499}{doi:\nolinkurl{10.18653/v1/2024.emnlp-main.499}}


\bibitem[Thakur et~al\mbox{.}(2021)]%
        {thakur2021beir}
\bibfield{author}{\bibinfo{person}{Nandan Thakur}, \bibinfo{person}{Nils Reimers}, \bibinfo{person}{Andreas R{\"u}ckl{\'e}}, \bibinfo{person}{Abhishek Srivastava}, {and} \bibinfo{person}{Iryna Gurevych}.} \bibinfo{year}{2021}\natexlab{}.
\newblock \showarticletitle{{BEIR}: A Heterogeneous Benchmark for Zero-shot Evaluation of Information Retrieval Models}. In \bibinfo{booktitle}{\emph{Thirty-fifth Conference on Neural Information Processing Systems Datasets and Benchmarks Track (Round 2)}}.
\newblock
\urldef\tempurl%
\url{https://openreview.net/forum?id=wCu6T5xFjeJ}
\showURL{%
\tempurl}


\bibitem[Thorne et~al\mbox{.}(2018a)]%
        {thorne-etal-2018-fever}
\bibfield{author}{\bibinfo{person}{James Thorne}, \bibinfo{person}{Andreas Vlachos}, \bibinfo{person}{Christos Christodoulopoulos}, {and} \bibinfo{person}{Arpit Mittal}.} \bibinfo{year}{2018}\natexlab{a}.
\newblock \showarticletitle{{FEVER}: a Large-scale Dataset for Fact Extraction and {VER}ification}. In \bibinfo{booktitle}{\emph{Proceedings of the 2018 Conference of the North {A}merican Chapter of the Association for Computational Linguistics: Human Language Technologies, Volume 1 (Long Papers)}}, \bibfield{editor}{\bibinfo{person}{Marilyn Walker}, \bibinfo{person}{Heng Ji}, {and} \bibinfo{person}{Amanda Stent}} (Eds.). \bibinfo{publisher}{Association for Computational Linguistics}, \bibinfo{address}{New Orleans, Louisiana}, \bibinfo{pages}{809--819}.
\newblock
\href{https://doi.org/10.18653/v1/N18-1074}{doi:\nolinkurl{10.18653/v1/N18-1074}}


\bibitem[Thorne et~al\mbox{.}(2018b)]%
        {thorne-etal-2018-fact}
\bibfield{author}{\bibinfo{person}{James Thorne}, \bibinfo{person}{Andreas Vlachos}, \bibinfo{person}{Oana Cocarascu}, \bibinfo{person}{Christos Christodoulopoulos}, {and} \bibinfo{person}{Arpit Mittal}.} \bibinfo{year}{2018}\natexlab{b}.
\newblock \showarticletitle{The Fact Extraction and {VER}ification ({FEVER}) Shared Task}. In \bibinfo{booktitle}{\emph{Proceedings of the First Workshop on Fact Extraction and {VER}ification ({FEVER})}}, \bibfield{editor}{\bibinfo{person}{James Thorne}, \bibinfo{person}{Andreas Vlachos}, \bibinfo{person}{Oana Cocarascu}, \bibinfo{person}{Christos Christodoulopoulos}, {and} \bibinfo{person}{Arpit Mittal}} (Eds.). \bibinfo{publisher}{Association for Computational Linguistics}, \bibinfo{address}{Brussels, Belgium}, \bibinfo{pages}{1--9}.
\newblock
\href{https://doi.org/10.18653/v1/W18-5501}{doi:\nolinkurl{10.18653/v1/W18-5501}}


\bibitem[Tian et~al\mbox{.}(2025)]%
        {10.1007/978-3-031-88708-6_3}
\bibfield{author}{\bibinfo{person}{Fangzheng Tian}, \bibinfo{person}{Debasis Ganguly}, {and} \bibinfo{person}{Craig Macdonald}.} \bibinfo{year}{2025}\natexlab{}.
\newblock \showarticletitle{Is Relevance Propagated from Retriever to Generator in RAG?}. In \bibinfo{booktitle}{\emph{Advances in Information Retrieval: 47th European Conference on Information Retrieval, ECIR 2025, Lucca, Italy, April 6–10, 2025, Proceedings, Part I}} (Lucca, Italy). \bibinfo{publisher}{Springer-Verlag}, \bibinfo{address}{Berlin, Heidelberg}, \bibinfo{pages}{32–48}.
\newblock
\showISBNx{978-3-031-88707-9}
\href{https://doi.org/10.1007/978-3-031-88708-6_3}{doi:\nolinkurl{10.1007/978-3-031-88708-6_3}}


\bibitem[Tonglet et~al\mbox{.}(2024)]%
        {tonglet-etal-2024-image}
\bibfield{author}{\bibinfo{person}{Jonathan Tonglet}, \bibinfo{person}{Marie-Francine Moens}, {and} \bibinfo{person}{Iryna Gurevych}.} \bibinfo{year}{2024}\natexlab{}.
\newblock \showarticletitle{{``}Image, Tell me your story!{''} Predicting the original meta-context of visual misinformation}. In \bibinfo{booktitle}{\emph{Proceedings of the 2024 Conference on Empirical Methods in Natural Language Processing}}, \bibfield{editor}{\bibinfo{person}{Yaser Al-Onaizan}, \bibinfo{person}{Mohit Bansal}, {and} \bibinfo{person}{Yun-Nung Chen}} (Eds.). \bibinfo{publisher}{Association for Computational Linguistics}, \bibinfo{address}{Miami, Florida, USA}, \bibinfo{pages}{7845--7864}.
\newblock
\href{https://doi.org/10.18653/v1/2024.emnlp-main.448}{doi:\nolinkurl{10.18653/v1/2024.emnlp-main.448}}


\bibitem[Ullrich et~al\mbox{.}(2024)]%
        {ullrich-etal-2024-aic}
\bibfield{author}{\bibinfo{person}{Herbert Ullrich}, \bibinfo{person}{Tom{\'a}{\v{s}} Mlyn{\'a}{\v{r}}}, {and} \bibinfo{person}{Jan Drchal}.} \bibinfo{year}{2024}\natexlab{}.
\newblock \showarticletitle{{AIC} {CTU} system at {AV}eri{T}e{C}: Re-framing automated fact-checking as a simple {RAG} task}. In \bibinfo{booktitle}{\emph{Proceedings of the Seventh Fact Extraction and VERification Workshop (FEVER)}}, \bibfield{editor}{\bibinfo{person}{Michael Schlichtkrull}, \bibinfo{person}{Yulong Chen}, \bibinfo{person}{Chenxi Whitehouse}, \bibinfo{person}{Zhenyun Deng}, \bibinfo{person}{Mubashara Akhtar}, \bibinfo{person}{Rami Aly}, \bibinfo{person}{Zhijiang Guo}, \bibinfo{person}{Christos Christodoulopoulos}, \bibinfo{person}{Oana Cocarascu}, \bibinfo{person}{Arpit Mittal}, \bibinfo{person}{James Thorne}, {and} \bibinfo{person}{Andreas Vlachos}} (Eds.). \bibinfo{publisher}{Association for Computational Linguistics}, \bibinfo{address}{Miami, Florida, USA}, \bibinfo{pages}{137--150}.
\newblock
\href{https://doi.org/10.18653/v1/2024.fever-1.16}{doi:\nolinkurl{10.18653/v1/2024.fever-1.16}}


\bibitem[Van~Landeghem et~al\mbox{.}(2023)]%
        {van2023document}
\bibfield{author}{\bibinfo{person}{Jordy Van~Landeghem}, \bibinfo{person}{Rub{\`e}n Tito}, \bibinfo{person}{{\L}ukasz Borchmann}, \bibinfo{person}{Micha{\l} Pietruszka}, \bibinfo{person}{Pawel Joziak}, \bibinfo{person}{Rafal Powalski}, \bibinfo{person}{Dawid Jurkiewicz}, \bibinfo{person}{Micka{\"e}l Coustaty}, \bibinfo{person}{Bertrand Anckaert}, \bibinfo{person}{Ernest Valveny}, {et~al\mbox{.}}} \bibinfo{year}{2023}\natexlab{}.
\newblock \showarticletitle{Document understanding dataset and evaluation (dude)}. In \bibinfo{booktitle}{\emph{Proceedings of the IEEE/CVF International Conference on Computer Vision}}. \bibinfo{pages}{19528--19540}.
\newblock


\bibitem[Viswanathan et~al\mbox{.}(2021)]%
        {viswanathan-etal-2021-citationie}
\bibfield{author}{\bibinfo{person}{Vijay Viswanathan}, \bibinfo{person}{Graham Neubig}, {and} \bibinfo{person}{Pengfei Liu}.} \bibinfo{year}{2021}\natexlab{}.
\newblock \showarticletitle{{C}itation{IE}: Leveraging the Citation Graph for Scientific Information Extraction}. In \bibinfo{booktitle}{\emph{Proceedings of the 59th Annual Meeting of the Association for Computational Linguistics and the 11th International Joint Conference on Natural Language Processing (Volume 1: Long Papers)}}, \bibfield{editor}{\bibinfo{person}{Chengqing Zong}, \bibinfo{person}{Fei Xia}, \bibinfo{person}{Wenjie Li}, {and} \bibinfo{person}{Roberto Navigli}} (Eds.). \bibinfo{publisher}{Association for Computational Linguistics}, \bibinfo{address}{Online}, \bibinfo{pages}{719--731}.
\newblock
\href{https://doi.org/10.18653/v1/2021.acl-long.59}{doi:\nolinkurl{10.18653/v1/2021.acl-long.59}}


\bibitem[Vladika and Matthes(2023)]%
        {vladika-matthes-2023-scientific}
\bibfield{author}{\bibinfo{person}{Juraj Vladika} {and} \bibinfo{person}{Florian Matthes}.} \bibinfo{year}{2023}\natexlab{}.
\newblock \showarticletitle{Scientific Fact-Checking: A Survey of Resources and Approaches}. In \bibinfo{booktitle}{\emph{Findings of the Association for Computational Linguistics: ACL 2023}}, \bibfield{editor}{\bibinfo{person}{Anna Rogers}, \bibinfo{person}{Jordan Boyd-Graber}, {and} \bibinfo{person}{Naoaki Okazaki}} (Eds.). \bibinfo{publisher}{Association for Computational Linguistics}, \bibinfo{address}{Toronto, Canada}, \bibinfo{pages}{6215--6230}.
\newblock
\href{https://doi.org/10.18653/v1/2023.findings-acl.387}{doi:\nolinkurl{10.18653/v1/2023.findings-acl.387}}


\bibitem[Vladika and Matthes(2024a)]%
        {vladika-matthes-2024-comparing}
\bibfield{author}{\bibinfo{person}{Juraj Vladika} {and} \bibinfo{person}{Florian Matthes}.} \bibinfo{year}{2024}\natexlab{a}.
\newblock \showarticletitle{Comparing Knowledge Sources for Open-Domain Scientific Claim Verification}. In \bibinfo{booktitle}{\emph{Proceedings of the 18th Conference of the European Chapter of the Association for Computational Linguistics (Volume 1: Long Papers)}}, \bibfield{editor}{\bibinfo{person}{Yvette Graham} {and} \bibinfo{person}{Matthew Purver}} (Eds.). \bibinfo{publisher}{Association for Computational Linguistics}, \bibinfo{address}{St. Julian{'}s, Malta}, \bibinfo{pages}{2103--2114}.
\newblock
\urldef\tempurl%
\url{https://aclanthology.org/2024.eacl-long.128/}
\showURL{%
\tempurl}


\bibitem[Vladika and Matthes(2024b)]%
        {vladika-matthes-2024-improving}
\bibfield{author}{\bibinfo{person}{Juraj Vladika} {and} \bibinfo{person}{Florian Matthes}.} \bibinfo{year}{2024}\natexlab{b}.
\newblock \showarticletitle{Improving Health Question Answering with Reliable and Time-Aware Evidence Retrieval}. In \bibinfo{booktitle}{\emph{Findings of the Association for Computational Linguistics: NAACL 2024}}, \bibfield{editor}{\bibinfo{person}{Kevin Duh}, \bibinfo{person}{Helena Gomez}, {and} \bibinfo{person}{Steven Bethard}} (Eds.). \bibinfo{publisher}{Association for Computational Linguistics}, \bibinfo{address}{Mexico City, Mexico}, \bibinfo{pages}{4752--4763}.
\newblock
\href{https://doi.org/10.18653/v1/2024.findings-naacl.295}{doi:\nolinkurl{10.18653/v1/2024.findings-naacl.295}}


\bibitem[Wadden et~al\mbox{.}(2020)]%
        {wadden-etal-2020-fact}
\bibfield{author}{\bibinfo{person}{David Wadden}, \bibinfo{person}{Shanchuan Lin}, \bibinfo{person}{Kyle Lo}, \bibinfo{person}{Lucy~Lu Wang}, \bibinfo{person}{Madeleine van Zuylen}, \bibinfo{person}{Arman Cohan}, {and} \bibinfo{person}{Hannaneh Hajishirzi}.} \bibinfo{year}{2020}\natexlab{}.
\newblock \showarticletitle{Fact or Fiction: Verifying Scientific Claims}. In \bibinfo{booktitle}{\emph{Proceedings of the 2020 Conference on Empirical Methods in Natural Language Processing (EMNLP)}}, \bibfield{editor}{\bibinfo{person}{Bonnie Webber}, \bibinfo{person}{Trevor Cohn}, \bibinfo{person}{Yulan He}, {and} \bibinfo{person}{Yang Liu}} (Eds.). \bibinfo{publisher}{Association for Computational Linguistics}, \bibinfo{address}{Online}, \bibinfo{pages}{7534--7550}.
\newblock
\href{https://doi.org/10.18653/v1/2020.emnlp-main.609}{doi:\nolinkurl{10.18653/v1/2020.emnlp-main.609}}


\bibitem[Wadden et~al\mbox{.}(2022a)]%
        {wadden-etal-2022-scifact}
\bibfield{author}{\bibinfo{person}{David Wadden}, \bibinfo{person}{Kyle Lo}, \bibinfo{person}{Bailey Kuehl}, \bibinfo{person}{Arman Cohan}, \bibinfo{person}{Iz Beltagy}, \bibinfo{person}{Lucy~Lu Wang}, {and} \bibinfo{person}{Hannaneh Hajishirzi}.} \bibinfo{year}{2022}\natexlab{a}.
\newblock \showarticletitle{{S}ci{F}act-Open: Towards open-domain scientific claim verification}. In \bibinfo{booktitle}{\emph{Findings of the Association for Computational Linguistics: EMNLP 2022}}, \bibfield{editor}{\bibinfo{person}{Yoav Goldberg}, \bibinfo{person}{Zornitsa Kozareva}, {and} \bibinfo{person}{Yue Zhang}} (Eds.). \bibinfo{publisher}{Association for Computational Linguistics}, \bibinfo{address}{Abu Dhabi, United Arab Emirates}, \bibinfo{pages}{4719--4734}.
\newblock
\href{https://doi.org/10.18653/v1/2022.findings-emnlp.347}{doi:\nolinkurl{10.18653/v1/2022.findings-emnlp.347}}


\bibitem[Wadden et~al\mbox{.}(2022b)]%
        {wadden-etal-2022-multivers}
\bibfield{author}{\bibinfo{person}{David Wadden}, \bibinfo{person}{Kyle Lo}, \bibinfo{person}{Lucy~Lu Wang}, \bibinfo{person}{Arman Cohan}, \bibinfo{person}{Iz Beltagy}, {and} \bibinfo{person}{Hannaneh Hajishirzi}.} \bibinfo{year}{2022}\natexlab{b}.
\newblock \showarticletitle{{M}ulti{V}er{S}: Improving scientific claim verification with weak supervision and full-document context}. In \bibinfo{booktitle}{\emph{Findings of the Association for Computational Linguistics: NAACL 2022}}, \bibfield{editor}{\bibinfo{person}{Marine Carpuat}, \bibinfo{person}{Marie-Catherine de~Marneffe}, {and} \bibinfo{person}{Ivan~Vladimir Meza~Ruiz}} (Eds.). \bibinfo{publisher}{Association for Computational Linguistics}, \bibinfo{address}{Seattle, United States}, \bibinfo{pages}{61--76}.
\newblock
\href{https://doi.org/10.18653/v1/2022.findings-naacl.6}{doi:\nolinkurl{10.18653/v1/2022.findings-naacl.6}}


\bibitem[Wang et~al\mbox{.}(2022)]%
        {wang2022mtcut}
\bibfield{author}{\bibinfo{person}{Dong Wang}, \bibinfo{person}{Jianxin Li}, \bibinfo{person}{Tianchen Zhu}, \bibinfo{person}{Haoyi Zhou}, \bibinfo{person}{Qishan Zhu}, \bibinfo{person}{Yuxin Wen}, {and} \bibinfo{person}{Hongming Piao}.} \bibinfo{year}{2022}\natexlab{}.
\newblock \showarticletitle{MtCut: A Multi-Task Framework for Ranked List Truncation}. In \bibinfo{booktitle}{\emph{Proceedings of the Fifteenth ACM International Conference on Web Search and Data Mining}}. \bibinfo{pages}{1054--1062}.
\newblock


\bibitem[Wang et~al\mbox{.}(2023b)]%
        {wang-etal-2023-check-covid}
\bibfield{author}{\bibinfo{person}{Gengyu Wang}, \bibinfo{person}{Kate Harwood}, \bibinfo{person}{Lawrence Chillrud}, \bibinfo{person}{Amith Ananthram}, \bibinfo{person}{Melanie Subbiah}, {and} \bibinfo{person}{Kathleen McKeown}.} \bibinfo{year}{2023}\natexlab{b}.
\newblock \showarticletitle{Check-{COVID}: Fact-Checking {COVID}-19 News Claims with Scientific Evidence}. In \bibinfo{booktitle}{\emph{Findings of the Association for Computational Linguistics: ACL 2023}}, \bibfield{editor}{\bibinfo{person}{Anna Rogers}, \bibinfo{person}{Jordan Boyd-Graber}, {and} \bibinfo{person}{Naoaki Okazaki}} (Eds.). \bibinfo{publisher}{Association for Computational Linguistics}, \bibinfo{address}{Toronto, Canada}, \bibinfo{pages}{14114--14127}.
\newblock
\href{https://doi.org/10.18653/v1/2023.findings-acl.888}{doi:\nolinkurl{10.18653/v1/2023.findings-acl.888}}


\bibitem[Wang et~al\mbox{.}(2024)]%
        {wang-etal-2024-factcheck}
\bibfield{author}{\bibinfo{person}{Yuxia Wang}, \bibinfo{person}{Revanth Gangi~Reddy}, \bibinfo{person}{Zain~Muhammad Mujahid}, \bibinfo{person}{Arnav Arora}, \bibinfo{person}{Aleksandr Rubashevskii}, \bibinfo{person}{Jiahui Geng}, \bibinfo{person}{Osama Mohammed~Afzal}, \bibinfo{person}{Liangming Pan}, \bibinfo{person}{Nadav Borenstein}, \bibinfo{person}{Aditya Pillai}, \bibinfo{person}{Isabelle Augenstein}, \bibinfo{person}{Iryna Gurevych}, {and} \bibinfo{person}{Preslav Nakov}.} \bibinfo{year}{2024}\natexlab{}.
\newblock \showarticletitle{Factcheck-Bench: Fine-Grained Evaluation Benchmark for Automatic Fact-checkers}. In \bibinfo{booktitle}{\emph{Findings of the Association for Computational Linguistics: EMNLP 2024}}, \bibfield{editor}{\bibinfo{person}{Yaser Al-Onaizan}, \bibinfo{person}{Mohit Bansal}, {and} \bibinfo{person}{Yun-Nung Chen}} (Eds.). \bibinfo{publisher}{Association for Computational Linguistics}, \bibinfo{address}{Miami, Florida, USA}, \bibinfo{pages}{14199--14230}.
\newblock
\href{https://doi.org/10.18653/v1/2024.findings-emnlp.830}{doi:\nolinkurl{10.18653/v1/2024.findings-emnlp.830}}


\bibitem[Wang et~al\mbox{.}(2023a)]%
        {wang2023learning}
\bibfield{author}{\bibinfo{person}{Zhiruo Wang}, \bibinfo{person}{Jun Araki}, \bibinfo{person}{Zhengbao Jiang}, \bibinfo{person}{Md~Rizwan Parvez}, {and} \bibinfo{person}{Graham Neubig}.} \bibinfo{year}{2023}\natexlab{a}.
\newblock \showarticletitle{Learning to filter context for retrieval-augmented generation}.
\newblock \bibinfo{journal}{\emph{arXiv preprint arXiv:2311.08377}} (\bibinfo{year}{2023}).
\newblock


\bibitem[West and Bergstrom(2021)]%
        {west2021misinformation}
\bibfield{author}{\bibinfo{person}{Jevin~D West} {and} \bibinfo{person}{Carl~T Bergstrom}.} \bibinfo{year}{2021}\natexlab{}.
\newblock \showarticletitle{Misinformation in and about science}.
\newblock \bibinfo{journal}{\emph{Proceedings of the National Academy of Sciences}} \bibinfo{volume}{118}, \bibinfo{number}{15} (\bibinfo{year}{2021}), \bibinfo{pages}{e1912444117}.
\newblock


\bibitem[Wright and Augenstein(2021)]%
        {wright-augenstein-2021-citeworth}
\bibfield{author}{\bibinfo{person}{Dustin Wright} {and} \bibinfo{person}{Isabelle Augenstein}.} \bibinfo{year}{2021}\natexlab{}.
\newblock \showarticletitle{{C}ite{W}orth: Cite-Worthiness Detection for Improved Scientific Document Understanding}. In \bibinfo{booktitle}{\emph{Findings of the Association for Computational Linguistics: ACL-IJCNLP 2021}}, \bibfield{editor}{\bibinfo{person}{Chengqing Zong}, \bibinfo{person}{Fei Xia}, \bibinfo{person}{Wenjie Li}, {and} \bibinfo{person}{Roberto Navigli}} (Eds.). \bibinfo{publisher}{Association for Computational Linguistics}, \bibinfo{address}{Online}, \bibinfo{pages}{1796--1807}.
\newblock
\href{https://doi.org/10.18653/v1/2021.findings-acl.157}{doi:\nolinkurl{10.18653/v1/2021.findings-acl.157}}


\bibitem[Wu et~al\mbox{.}(2021)]%
        {wu2021learning}
\bibfield{author}{\bibinfo{person}{Chen Wu}, \bibinfo{person}{Ruqing Zhang}, \bibinfo{person}{Jiafeng Guo}, \bibinfo{person}{Yixing Fan}, \bibinfo{person}{Yanyan Lan}, {and} \bibinfo{person}{Xueqi Cheng}.} \bibinfo{year}{2021}\natexlab{}.
\newblock \showarticletitle{Learning to truncate ranked lists for information retrieval}. In \bibinfo{booktitle}{\emph{Proceedings of the AAAI Conference on Artificial Intelligence}}, Vol.~\bibinfo{volume}{35}. \bibinfo{pages}{4453--4461}.
\newblock


\bibitem[Wu et~al\mbox{.}(2024)]%
        {wu-etal-2024-scimmir}
\bibfield{author}{\bibinfo{person}{Siwei Wu}, \bibinfo{person}{Yizhi Li}, \bibinfo{person}{Kang Zhu}, \bibinfo{person}{Ge Zhang}, \bibinfo{person}{Yiming Liang}, \bibinfo{person}{Kaijing Ma}, \bibinfo{person}{Chenghao Xiao}, \bibinfo{person}{Haoran Zhang}, \bibinfo{person}{Bohao Yang}, \bibinfo{person}{Wenhu Chen}, \bibinfo{person}{Wenhao Huang}, \bibinfo{person}{Noura Al~Moubayed}, \bibinfo{person}{Jie Fu}, {and} \bibinfo{person}{Chenghua Lin}.} \bibinfo{year}{2024}\natexlab{}.
\newblock \showarticletitle{{S}ci{MMIR}: Benchmarking Scientific Multi-modal Information Retrieval}. In \bibinfo{booktitle}{\emph{Findings of the Association for Computational Linguistics: ACL 2024}}, \bibfield{editor}{\bibinfo{person}{Lun-Wei Ku}, \bibinfo{person}{Andre Martins}, {and} \bibinfo{person}{Vivek Srikumar}} (Eds.). \bibinfo{publisher}{Association for Computational Linguistics}, \bibinfo{address}{Bangkok, Thailand}, \bibinfo{pages}{12560--12574}.
\newblock
\href{https://doi.org/10.18653/v1/2024.findings-acl.746}{doi:\nolinkurl{10.18653/v1/2024.findings-acl.746}}


\bibitem[W{\"u}hrl and Klinger(2022)]%
        {wuhrl-klinger-2022-entity}
\bibfield{author}{\bibinfo{person}{Amelie W{\"u}hrl} {and} \bibinfo{person}{Roman Klinger}.} \bibinfo{year}{2022}\natexlab{}.
\newblock \showarticletitle{Entity-based Claim Representation Improves Fact-Checking of Medical Content in Tweets}. In \bibinfo{booktitle}{\emph{Proceedings of the 9th Workshop on Argument Mining}}, \bibfield{editor}{\bibinfo{person}{Gabriella Lapesa}, \bibinfo{person}{Jodi Schneider}, \bibinfo{person}{Yohan Jo}, {and} \bibinfo{person}{Sougata Saha}} (Eds.). \bibinfo{publisher}{International Conference on Computational Linguistics}, \bibinfo{address}{Online and in Gyeongju, Republic of Korea}, \bibinfo{pages}{187--198}.
\newblock
\urldef\tempurl%
\url{https://aclanthology.org/2022.argmining-1.18/}
\showURL{%
\tempurl}


\bibitem[Yao et~al\mbox{.}(2023)]%
        {yao2023end}
\bibfield{author}{\bibinfo{person}{Barry~Menglong Yao}, \bibinfo{person}{Aditya Shah}, \bibinfo{person}{Lichao Sun}, \bibinfo{person}{Jin-Hee Cho}, {and} \bibinfo{person}{Lifu Huang}.} \bibinfo{year}{2023}\natexlab{}.
\newblock \showarticletitle{End-to-end multimodal fact-checking and explanation generation: A challenging dataset and models}. In \bibinfo{booktitle}{\emph{Proceedings of the 46th International ACM SIGIR Conference on Research and Development in Information Retrieval}}. \bibinfo{pages}{2733--2743}.
\newblock


\bibitem[Ye et~al\mbox{.}(2023)]%
        {ye2023large}
\bibfield{author}{\bibinfo{person}{Yunhu Ye}, \bibinfo{person}{Binyuan Hui}, \bibinfo{person}{Min Yang}, \bibinfo{person}{Binhua Li}, \bibinfo{person}{Fei Huang}, {and} \bibinfo{person}{Yongbin Li}.} \bibinfo{year}{2023}\natexlab{}.
\newblock \showarticletitle{Large language models are versatile decomposers: Decomposing evidence and questions for table-based reasoning}. In \bibinfo{booktitle}{\emph{Proceedings of the 46th International ACM SIGIR Conference on Research and Development in Information Retrieval}}. \bibinfo{pages}{174--184}.
\newblock


\bibitem[Yin et~al\mbox{.}(2020)]%
        {yin-etal-2020-tabert}
\bibfield{author}{\bibinfo{person}{Pengcheng Yin}, \bibinfo{person}{Graham Neubig}, \bibinfo{person}{Wen-tau Yih}, {and} \bibinfo{person}{Sebastian Riedel}.} \bibinfo{year}{2020}\natexlab{}.
\newblock \showarticletitle{{T}a{BERT}: Pretraining for Joint Understanding of Textual and Tabular Data}. In \bibinfo{booktitle}{\emph{Proceedings of the 58th Annual Meeting of the Association for Computational Linguistics}}, \bibfield{editor}{\bibinfo{person}{Dan Jurafsky}, \bibinfo{person}{Joyce Chai}, \bibinfo{person}{Natalie Schluter}, {and} \bibinfo{person}{Joel Tetreault}} (Eds.). \bibinfo{publisher}{Association for Computational Linguistics}, \bibinfo{address}{Online}, \bibinfo{pages}{8413--8426}.
\newblock
\href{https://doi.org/10.18653/v1/2020.acl-main.745}{doi:\nolinkurl{10.18653/v1/2020.acl-main.745}}


\bibitem[Zamani and Bendersky(2024)]%
        {zamani2024stochastic}
\bibfield{author}{\bibinfo{person}{Hamed Zamani} {and} \bibinfo{person}{Michael Bendersky}.} \bibinfo{year}{2024}\natexlab{}.
\newblock \showarticletitle{Stochastic rag: End-to-end retrieval-augmented generation through expected utility maximization}. In \bibinfo{booktitle}{\emph{Proceedings of the 47th International ACM SIGIR Conference on Research and Development in Information Retrieval}}. \bibinfo{pages}{2641--2646}.
\newblock


\bibitem[Zeng et~al\mbox{.}(2021)]%
        {zeng2021automated}
\bibfield{author}{\bibinfo{person}{Xia Zeng}, \bibinfo{person}{Amani~S Abumansour}, {and} \bibinfo{person}{Arkaitz Zubiaga}.} \bibinfo{year}{2021}\natexlab{}.
\newblock \showarticletitle{Automated fact-checking: A survey}.
\newblock \bibinfo{journal}{\emph{Language and Linguistics Compass}} \bibinfo{volume}{15}, \bibinfo{number}{10} (\bibinfo{year}{2021}), \bibinfo{pages}{e12438}.
\newblock


\bibitem[Zhang et~al\mbox{.}(2023)]%
        {zhang-etal-2023-relevance}
\bibfield{author}{\bibinfo{person}{Hengran Zhang}, \bibinfo{person}{Ruqing Zhang}, \bibinfo{person}{Jiafeng Guo}, \bibinfo{person}{Maarten de Rijke}, \bibinfo{person}{Yixing Fan}, {and} \bibinfo{person}{Xueqi Cheng}.} \bibinfo{year}{2023}\natexlab{}.
\newblock \showarticletitle{From Relevance to Utility: Evidence Retrieval with Feedback for Fact Verification}. In \bibinfo{booktitle}{\emph{Findings of the Association for Computational Linguistics: EMNLP 2023}}, \bibfield{editor}{\bibinfo{person}{Houda Bouamor}, \bibinfo{person}{Juan Pino}, {and} \bibinfo{person}{Kalika Bali}} (Eds.). \bibinfo{publisher}{Association for Computational Linguistics}, \bibinfo{address}{Singapore}, \bibinfo{pages}{6373--6384}.
\newblock
\href{https://doi.org/10.18653/v1/2023.findings-emnlp.422}{doi:\nolinkurl{10.18653/v1/2023.findings-emnlp.422}}


\bibitem[Zhang and Abernethy(2024)]%
        {zhang2024detecting}
\bibfield{author}{\bibinfo{person}{Tianmai~M Zhang} {and} \bibinfo{person}{Neil~F Abernethy}.} \bibinfo{year}{2024}\natexlab{}.
\newblock \showarticletitle{Detecting Reference Errors in Scientific Literature with Large Language Models}.
\newblock \bibinfo{journal}{\emph{arXiv preprint arXiv:2411.06101}} (\bibinfo{year}{2024}).
\newblock


\bibitem[Zhang et~al\mbox{.}(2021)]%
        {zhang-etal-2021-abstract}
\bibfield{author}{\bibinfo{person}{Zhiwei Zhang}, \bibinfo{person}{Jiyi Li}, \bibinfo{person}{Fumiyo Fukumoto}, {and} \bibinfo{person}{Yanming Ye}.} \bibinfo{year}{2021}\natexlab{}.
\newblock \showarticletitle{Abstract, Rationale, Stance: A Joint Model for Scientific Claim Verification}. In \bibinfo{booktitle}{\emph{Proceedings of the 2021 Conference on Empirical Methods in Natural Language Processing}}, \bibfield{editor}{\bibinfo{person}{Marie-Francine Moens}, \bibinfo{person}{Xuanjing Huang}, \bibinfo{person}{Lucia Specia}, {and} \bibinfo{person}{Scott Wen-tau Yih}} (Eds.). \bibinfo{publisher}{Association for Computational Linguistics}, \bibinfo{address}{Online and Punta Cana, Dominican Republic}, \bibinfo{pages}{3580--3586}.
\newblock
\href{https://doi.org/10.18653/v1/2021.emnlp-main.290}{doi:\nolinkurl{10.18653/v1/2021.emnlp-main.290}}


\bibitem[Zheng et~al\mbox{.}(2024)]%
        {zheng-etal-2024-evidence}
\bibfield{author}{\bibinfo{person}{Liwen Zheng}, \bibinfo{person}{Chaozhuo Li}, \bibinfo{person}{Xi Zhang}, \bibinfo{person}{Yu-Ming Shang}, \bibinfo{person}{Feiran Huang}, {and} \bibinfo{person}{Haoran Jia}.} \bibinfo{year}{2024}\natexlab{}.
\newblock \showarticletitle{Evidence Retrieval is almost All You Need for Fact Verification}. In \bibinfo{booktitle}{\emph{Findings of the Association for Computational Linguistics ACL 2024}}, \bibfield{editor}{\bibinfo{person}{Lun-Wei Ku}, \bibinfo{person}{Andre Martins}, {and} \bibinfo{person}{Vivek Srikumar}} (Eds.). \bibinfo{publisher}{Association for Computational Linguistics}, \bibinfo{address}{Bangkok, Thailand and virtual meeting}, \bibinfo{pages}{9274--9281}.
\newblock
\urldef\tempurl%
\url{https://aclanthology.org/2024.findings-acl.551}
\showURL{%
\tempurl}


\end{thebibliography}

\end{document}